\documentclass[aps,prl,twocolumn,superscriptaddress,longbibliography]{revtex4-2}
\usepackage{amsmath,amssymb}
\usepackage[pdftex]{hyperref,graphicx}
\hypersetup{colorlinks = true, urlcolor = blue, linkcolor = blue, citecolor = blue}
\usepackage{physics}
\usepackage{xcolor}
\usepackage{bm}

\begin{document}
\title{Topological Phases in AB-Stacked MoTe$_2$/WSe$_2$: $\mathbb{Z}_2$ Topological Insulators, Chern Insulators, and Topological Charge Density Waves}

\author{Haining Pan}
\affiliation{Condensed Matter Theory Center and Joint Quantum Institute, Department of Physics, University of Maryland, College Park, Maryland 20742, USA}

\author{Ming Xie}
\affiliation{Condensed Matter Theory Center and Joint Quantum Institute, Department of Physics, University of Maryland, College Park, Maryland 20742, USA}

\author{Fengcheng Wu}
\email{wufcheng@whu.edu.cn}
\affiliation{School of Physics and Technology, Wuhan University, Wuhan 430072, China}
\affiliation{Wuhan Institute of Quantum Technology, Wuhan 430206, China}

\author{Sankar Das Sarma}
\affiliation{Condensed Matter Theory Center and Joint Quantum Institute, Department of Physics, University of Maryland, College Park, Maryland 20742, USA}

\begin{abstract}
    We present a theory on the quantum phase diagram of AB-stacked MoTe$_2$/WSe$_2$ using a self-consistent Hartree-Fock calculation performed in the plane-wave basis, motivated by the observation of topological states in this system. At filling factor $\nu=2$ (two holes per moir\'e unit cell), Coulomb interaction can stabilize a $\mathbb{Z}_2$ topological insulator by opening a charge gap. At $\nu=1$, the interaction induces three classes of competing states, spin density wave states, an in-plane ferromagnetic state, and a valley polarized state, which undergo first-order phase transitions tuned by an out-of-plane displacement field. The valley polarized state becomes a Chern insulator for certain displacement fields.  Moreover, we predict a topological charge density wave forming a honeycomb lattice with ferromagnetism at $\nu=2/3$. Future directions on this versatile system hosting a rich set of quantum phases are discussed.
\end{abstract}

\maketitle

\textit{Introduction.---} Topological phases of matter have been among the most active and important research areas in condensed matter physics ever since the experimental discovery of the quantum Hall effect~\cite{vonklitzing1980new} in the presence of strong magnetic fields.
The mechanism of the quantum Hall effect, especially the quantization of Hall conductance, has been understood in terms of the topological invariant of the Chern number~\cite{thouless1982quantized,niu1985quantized}.
The concept of quantum Hall insulators has been generalized to quantum anomalous Hall insulators (also known as Chern insulators) \cite{haldane1988model} and quantum spin Hall insulators (also known as $\mathbb{Z}_2$ topological insulators) \cite{kane2005quantum}, where the former exhibits quantized Hall effect without the external magnetic field, and the latter is the topological state in the presence of time-reversal symmetry.
Among the extensive topological theoretical predictions~\cite{onoda2003quantized,qi2006topological,qi2008topological,liu2008quantum,yu2010quantized,qiao2010quantum,nomura2011surfacequantized,zhang2019topological,li2019intrinsic,otrokov2019prediction,liu2008quantuma,zhang2019nearly,bernevig2006quantum,sheng2006quantum,qian2014quantum}, only a limited number of systems have so far been studied experimentally manifesting unambiguous topological invariants: Chern insulators have been realized in magnetically doped topological insulator thin films~\cite{chang2013experimental,zhao2020tuning}, few-layer MnBi$_2$Te$_4$~\cite{deng2020quantum,liu2020robust}, and graphene-based moir\'e materials~\cite{sharpe2019emergent,serlin2020intrinsic,chen2020tunable,polshyn2020electrical,polshyn2022topological}, and transport signatures of the quantum spin Hall effect have been reported in HgTe quantum wells~\cite{konig2007quantum}, InAs/GaSb quantum wells~\cite{knez2010finite,knez2011evidence} and monolayer WTe$_2$~\cite{fei2017edge,wu2018observation}.

The recent advent of moir\'e materials, followed by the discovery of correlated insulators and superconductors in magic-angle twisted bilayer graphene~\cite{cao2018correlated,cao2018unconventional}, provides vast new opportunities to design different phases of matter, including the topological phases discussed above~\cite{yankowitz2019tuning,chen2019evidence,cao2020tunable,chen2020electrically,wong2020cascade,serlin2020intrinsic,sharpe2019emergent,lu2019superconductors,shen2020correlated,choi2019electronic,xie2019spectroscopic,chen2019signatures,chen2020tunable,liu2020tunable}. It was theoretically proposed \cite{wu2019topological} that moir\'e bands in twisted transition metal dichalcogenide homobilayers can be mapped to the Kane-Mele model~\cite{wu2019topological,pan2020band,devakul2021magic}. Coulomb repulsion can further drive broken symmetry insulating states (e.g., Chern insulators) because the topological moir\'e bands have narrow bandwidth. This leads to interesting physics involving the interplay between band topology and strong correlations. 
A recent experiment on AB-stacked MoTe$_2$/WSe$_2$ \cite{li2021quantum} reported well-developed quantum anomalous Hall effect at filling factor $\nu=1$, and possible evidence for quantum spin Hall effect at $\nu=2$. This experiment is unprecedented, as it demonstrates experimentally that quantum anomalous Hall and quantum spin Hall effects could be realized in a single system, something that has never happened before as the two phenomena are quite distinct. It is also surprising, as it utilizes a heterobilayer instead of the homobilayer proposed in Ref.~\onlinecite{wu2019topological} for the manifestation of Transition Metal Dichalcogenides (TMD) topology. The surprising rich phenomena observed in MoTe$_2$/WSe$_2$ calls for detailed theoretical studies \cite{xie2022valleypolarized,zhang2021spintextured,devakul2021quantum}.

In this Letter, we present a topological theoretical study of interacting AB-stacked MoTe$_2$/WSe$_2$. The single-particle physics is described by a continuum moir\'e Hamiltonian that incorporates the topmost valence bands from both layers. An out-of-plane displacement field tunes the band offset between the two layers, and drives a topological phase transition for the first moir\'e valence band. The Coulomb interaction is treated using a self-consistent Hartree-Fock approximation {\it without} the bias of projecting it to a few selected moir\'e bands. Our main results are summarized as follows.
{(1) At $\nu=2$, the single-particle band structure predicts a metallic state in the topological regime, but the Coulomb interaction can open up a gap, and therefore, stabilize the $\mathbb{Z}_2$ topological insulator. 
(2) At $\nu=1$, we obtain a rich quantum phase diagram that includes three classes of competing phases, spin density wave states, an in-plane ferromagnetic state, and a valley polarized state. The valley polarized state becomes a Chern insulator for a certain range of displacement fields. 
(3) At the fractional filling factor $\nu=2/3$, we predict a topological density wave state, and identify the condition for its experimental realization.

Our results provide a consistent mean-field description of the experimental observations (along with new predictions) in Ref.~\onlinecite{li2021quantum}, which is an important step toward a full understanding of quantum phases in the moir\'e TMD physics. In a broader perspective, our work establishes affirmatively that the interplay between many-body interaction and single-particle band topology can induce a rich set of distinct topological phases (e.g., quantum anomalous Hall insulators and quantum spin Hall insulators) within one realistic system.}

\begin{figure}[t]
    \centering
    \includegraphics[width=3.4in]{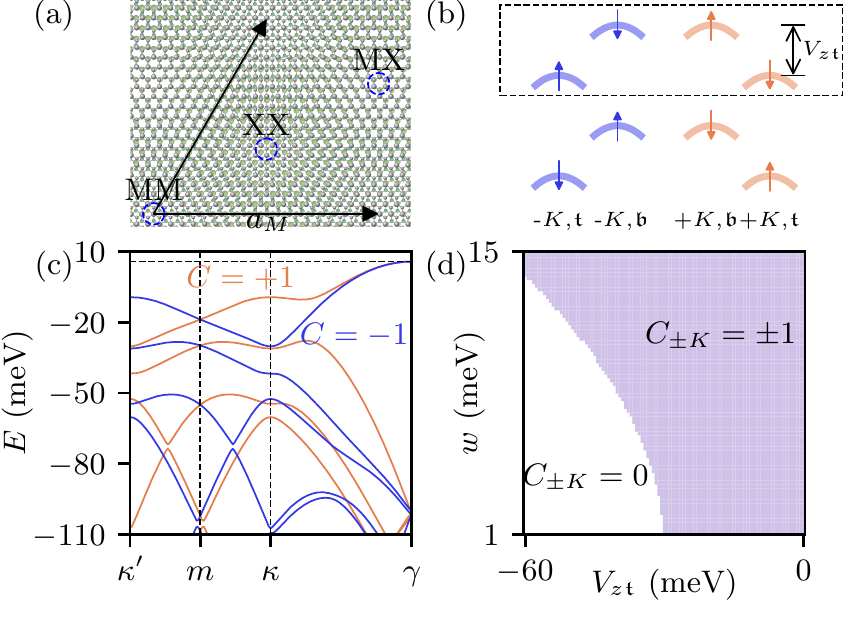}
    \caption{(a) Moir\'e superlattice of AB-stacked MoTe$_2$/WSe$_2$. (b) A schematic plot for band alignment in the heterobilayer. Only states in the dashed box are retained in the Hamiltonian $H_\tau$. (c) Single-particle moir\'e band structure with parameters ($w$, $V_\mathfrak{b}$, $V_{z\mathfrak{t}}$) =(12 meV, 7 meV, $-$20 meV). Orange (blue) lines show $+K$($-K$)-valley moir\'e bands. {The dashed line shows the Fermi energy of the charge neutrality.} The first moir\'e valence band from $\pm K$ valleys carry $\pm 1$ Chern number. {(See the Supplemental Material~\cite{HeteroBilayer_SM}\cite*{wu2020quantum, fukui2005chern, monkhorst1976special} for the definition of high-symmetry points in the moir\'e Brillouin zone.)} (d) Single-particle topological phase diagram of the first moir\'e valence band as a function of $w$ and $V_{z\mathfrak{t}}$ with $V_\mathfrak{b}$=7 meV.}
    \label{fig:1}
\end{figure}

\textit{Model.---}
We focus on AB-stacked MoTe$_2$/WSe$_2$ with an exact $180^\circ$ twist angle. The moir\'e superlattices have the $C_{3v}$ point group symmetry and a moir\'e period $a_M=a_{\mathfrak{b}}a_{\mathfrak{t}}/\abs{a_{\mathfrak{b}}-a_{\mathfrak{t}}}$, where $(a_{\mathfrak{b}},a_{\mathfrak{t}})$=(3.575\AA, 3.32\AA) are the lattice constant of MoTe$_2$ and WSe$_2$, respectively. The large lattice constant mismatch makes the moir\'e superlattices relatively immune to twist angle disorder. As shown in Fig.~\ref{fig:1} (a), there are three high-symmetry regions in one moir\'e unit cell, labeled as MM,  XX, and MX, corresponding to  local atomic configurations with $C_{3z}$ (threefold
rotational) symmetry.

The density functional theory calculation~\cite{zhang2021spintextured,li2021quantum} of AB-stacked MoTe$_2$/WSe$_2$ shows that states at the valence band edge originate from $\pm K$ valleys of MoTe$_2$, where $\pm K$ refer to corners of the monolayer Brillouin zone. Therefore, we focus on $\pm K$ valleys which are related by the $\mathcal{T}$ symmetry. For valence bands at the two valleys, there are large spin splittings that are both valley and layer dependent, as illustrated in Fig.~\ref{fig:1} (b).  By retaining the topmost valence bands from each layer, we construct \cite{wu2019topological,pan2020band,zhang2021spintextured} a valley-dependent continuum Hamiltonian as follows  
\begin{equation}\label{eq:Ham}
    H_{\tau}=\begin{pmatrix}
        -\frac{\hbar^2\bm{k}^2}{2m_\mathfrak{b}}+\Delta_{\mathfrak{b}}(\bm{r}) &  \Delta_{\text{T},\tau}(\bm{r})\\
        \Delta_{\text{T},\tau}^\dag(\bm{r}) & -\frac{\hbar^2\left(\bm{k}-\tau \bm{\kappa}\right)^2}{2m_\mathfrak{t}}+ \Delta_\mathfrak{t}(\bm{r})+V_{z\mathfrak{t}}
    \end{pmatrix},
\end{equation} 
where $\tau=\pm 1$ represents $\pm K$ valleys, and $\bm{\kappa}=\frac{4\pi}{3a_M}\left(1,0\right)$  is at a corner of the  moir\'e Brillouin zone.  For each valley, the $2\times 2$ Hamiltonian hybridizes the bottom layer ($\mathfrak{b}$) and top layer ($\mathfrak{t}$), where the off diagonal terms describe the interlayer tunneling $\Delta_{\text{T},\tau}$, and the diagonal terms describe the momentum-shifted kinetic energy with the effective mass $(m_{\mathfrak{b}},m_{\mathfrak{t}})=(0.65,0.35)m_e$ ($m_e$ is the rest electron mass), plus the intralayer potential $\Delta_{\mathfrak{b}/\mathfrak{t}}$, and a band offset $V_{z\mathfrak{t}}$ [Fig.~\ref{fig:1}(b)]. Our $H_\tau$ differs from that in Ref.~\onlinecite{zhang2021spintextured} by a gauge transformation. 

The periodic potential  $\Delta_{\mathfrak{b}}(\bm{r})$ is parametrized as 
\begin{equation}\label{eq:Delta_b}
    \Delta_{\mathfrak{b}}(\bm{r})=2V_{\mathfrak{b}}\sum_{j=1,3,5} \cos(\bm{g}_j \cdot \bm{r}+\psi_{\mathfrak{b}}),
\end{equation}
where $V_{\mathfrak{b}}$ and $\psi_{\mathfrak{b}}$ respectively characterize the amplitude and spatial pattern of the potential, and $\bm{g}_j=\frac{4\pi}{\sqrt{3} a_M} \left(- \sin \frac{\pi (j-1)}{3}, \cos \frac{\pi (j-1)}{3}\right)$ are the moir\'e reciprocal lattice vectors in the first shell. We set $\Delta_{\mathfrak{t}}(\bm{r})=0$, since the low-energy physics only involves the band maximum of WSe$_2$ \cite{zhang2021spintextured}.  The interlayer tunneling term is
\begin{equation}\label{eq:Delta_T}
    \Delta_{\text{T},\tau}(\bm{r})=\tau w \left(1+\omega^{\tau} e^{i\tau\bm{g}_2\cdot\bm{r}}+\omega^{2\tau} e^{i\tau\bm{g}_3\cdot\bm{r}} \right),
\end{equation}
where $w$ describes the tunneling strength, and $\omega=e^{i\frac{2\pi}{3}} $ following the $C_{3z}$ symmetry \cite{zhang2021spintextured}. The valley dependence of $ \Delta_{\text{T},\tau}$ is constrained by $\mathcal{T}$ symmetry. Here $ \Delta_{\text{T},\tau}$ couples states with opposite spins, which is symmetry allowed because the heterobilayer breaks the $z \rightarrow -z$  mirror symmetry. For parameters in $H_\tau$, we take $\psi_{\mathfrak{b}}=-14^{\circ}$ such that the potential maximum of $\Delta_{\mathfrak{b}}(\bm{r})$ is at the MM site \cite{zhang2021spintextured}; $V_{z\mathfrak{t}}$ is a parameter that is experimentally controllable by an applied out-of-plane displacement field; $V_{\mathfrak{b}}$ and $w$ are taken as theoretical parameters that can be adjusted to study different phases. We note that the interlayer tunneling 
strength $w$ can be modified by pressure. 

The low-energy moir\'e bands and their topology are tunable by $V_{z\mathfrak{t}}$. For the intrinsic band offset $V_{z\mathfrak{t}}$ that has a large negative value ($\sim -$ 110 meV) \cite{zhang2021spintextured}, the topmost moir\'e valence band at each valley mainly derives from the MoTe$_2$ layer and can be described by a tight-binding model on a triangular lattice, which is topologically trivial. By reducing $|V_{z\mathfrak{t}}|$ with an external displacement field, the energy gap between the topmost moir\'e valence bands derived respectively from MoTe$_2$ and WSe$_2$ can close and then reopen. This band inversion enabled by the tunneling term  $ \Delta_{\text{T},\tau}$ can lead to topological phase transitions. Figure~\ref{fig:1}(d) presents a topological phase diagram characterized by the valley-contrast Chern numbers $C_{\pm K}$ of the first moir\'e valence band in the parameter space of ($V_{z\mathfrak{t}}$, $w$). For $V_{z\mathfrak{t}}$ above the critical value, $C_{\pm K}$ become nontrivial with value ($\pm 1$). Figure~\ref{fig:1}(c) plots a representative moir\'e band structure in the topologically nontrivial regime of Fig.~\ref{fig:1}(d). Note that there is no overall energy gap that separates the first and second moir\'e valence bands in Fig.~\ref{fig:1}(c). The effect of many-body interaction on this gap is crucial as studied in the following.

\begin{figure}[t]
    \centering
    \includegraphics[width=3.4in]{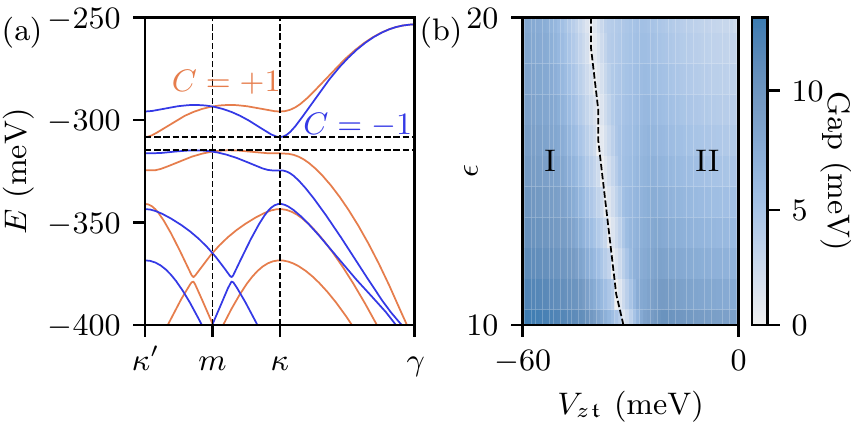}
    \caption{(a) Band structure from the HF calculation  at $\nu=2$ with the same parameters as Fig.~\ref{fig:1}(c). Orange (blue) lines show $+K$($-K$)-valley moir\'e bands. {The two horizontal dashed lines identify the charge gap at $\nu=2$.} Here $\epsilon=15$. (b) Charge gap at $\nu=2$ from the HF calculation as a function of $V_{z\mathfrak{t}}$ and $\epsilon$.  The gap vanishes at the dashed line, which separates phase I (topologically trivial insulators) and phase II ($\mathbb{Z}_2$ topological insulators). Here $w$=12 meV and $V_\mathfrak{b}$=7 meV.}
    \label{fig:2}
\end{figure}

\textit{Coulomb Interactions.---} The bandwidth of the first moir\'e valence band is on the order of $\hbar^2 \bm{\kappa}^2/(2 m_{\mathfrak{b}}) \approx 47 $ meV, while the characteristic Coulomb interaction strength is on the order of $e^2/(\epsilon a_M) \approx 31 $ meV for $a_M  \approx 4.7 $ nm and dielectric constant $\epsilon= 10 $. Therefore, the Coulomb interaction has an energy scale comparable to the bandwidth [Fig.~\ref{fig:1}(c)], which puts the system in the strongly interacting regime with possible interaction-induced quantum phase transitions. We consider the dual-gate screened Coulomb interaction with the momentum-dependent potential $V(q)=2\pi e^2 \tanh(q d)/(\epsilon q)$, where $d$ is the gate-to-sample distance. We set $d=5$ nm \cite{li2021quantum} unless otherwise stated. The dielectric constant $\epsilon$ is of the order of $10-20$ , and we vary $\epsilon$ to illustrate how physical properties depend on the interaction strength.

We treat the Coulomb interaction using a self-consistent Hartree-Fock (HF) approximation applied to the continuum Hamiltonian in the plane-wave basis (see the Supplemental Material ~\cite{HeteroBilayer_SM} for details). Here, we do {\it not} project the Coulomb interaction onto a few low-energy moir\'e bands, because it is generally not possible to identify a set of bands that are energetically isolated from other bands [Fig.~\ref{fig:1}(c)]. Our approach allows us to study interaction effects for different parameter regimes in a unified manner. We note that HF theory, although qualitatively reliable, may overestimate the tendency toward ordering, and may not be able to quantitatively capture the global phase diagram. Nevertheless, it is an important and nontrivial question of principle whether the experimental observations can be captured by a mean-field theory. 

\begin{figure}[t]
    \centering
    \includegraphics[width=3.4in]{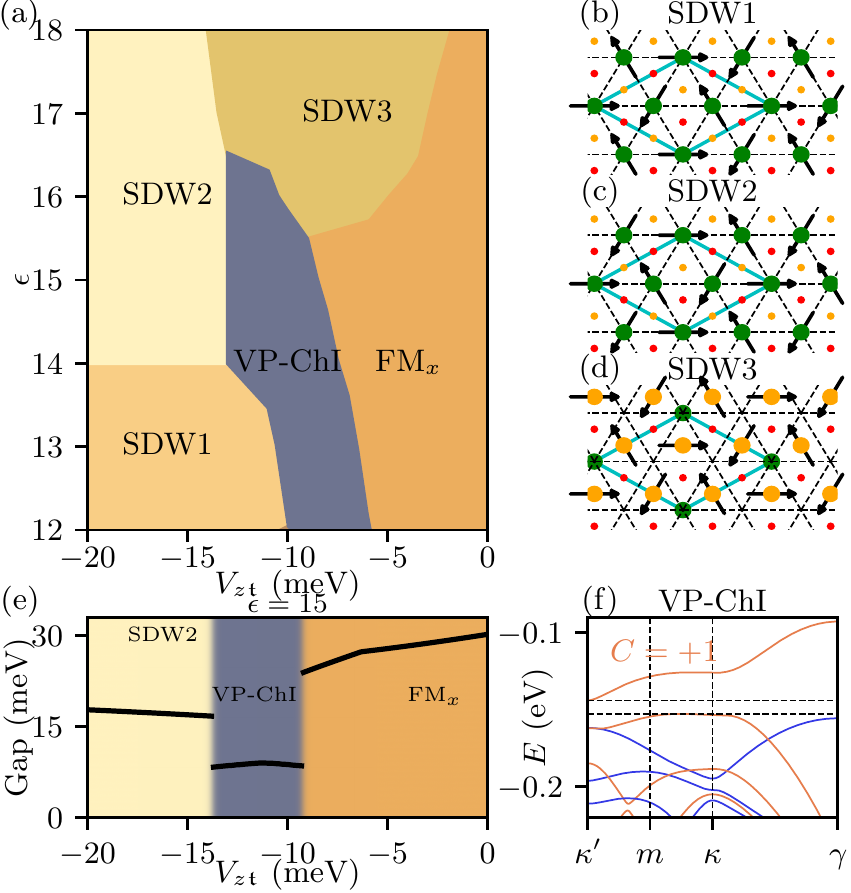}
    \caption{(a) Interaction-induced phase diagram at $\nu=1$ as a function of $V_{z\mathfrak{t}}$ and $\epsilon$.  (b)-(d) Spin textures in the bottom layer for the three SDW states. The green, orange and red dots mark MM, XX and {MX} sites, respectively, {where their sizes indicate the corresponding charge densities.} (e)  Charge gap at $\nu=1$ as a function of $V_{z\mathfrak{t}}$ with $\epsilon$=15, $w$=12 meV, and $V_\mathfrak{b}$=7 meV. (f) Band structure of $\nu=1$ VP ChI calculated using the HF method at $\epsilon=15$ and $V_{z\mathfrak{t}}=-10$ meV. Orange (blue) lines show $+K$($-K$)-valley moir\'e bands. {The two horizontal dashed lines identify the charge gap at $\nu=1$.}
        }
    \label{fig:3}
\end{figure}

\textit{$\mathbb{Z}_2$ topological insulator at $\nu=2$.---}
For the single-particle band structure shown in Fig.~\ref{fig:1}(c), $\nu=2$ would correspond to a metallic state since there is no overall gap that separates the first and second moir\'e valence bands. However, after performing the self-consistent HF calculations using the same set of parameters, we find that a true gap develops at $\nu=2$, as shown in  Fig.~\ref{fig:2}(a).  The first moir\'e valence band at each valley remains topologically nontrivial. Therefore, the $\nu=2$ insulator in  Fig.~\ref{fig:2}(a) belongs to
the $\mathbb{Z}_2$ time-reversal invariant topological insulator as the $\mathbb{Z}_2$ topological invariant, defined in our case by $\text{mod}[(C_{+K}-C_{-K})/2,2]$~\cite{kane2005quantum}, is nontrivial given $C_{\pm K}=\pm 1$. This can be viewed as a realization of the interaction-induced quantum spin Hall insulator~\cite{kane2005quantum}.

For generic parameter values, we present the topological phase diagram at $\nu=2$ calculated using the HF approximation as a function of $\epsilon$ and $V_{z\mathfrak{t}}$ in Fig.~\ref{fig:2}(b), where there are two phases, the topologically trivial phase and the $\mathbb{Z}_2 $ topological insulator phase, respectively for $V_{z\mathfrak{t}}$ below and above a critical value. At the critical $V_{z\mathfrak{t}}$, the charge gap vanishes.  The critical $V_{z\mathfrak{t}}$ increases as $\epsilon$ decreases (Coulomb interaction increases) because a stronger Coulomb interaction opens a larger charge gap in the topologically trivial phase, which requires a larger displacement field to close the gap so that the topological phase transition can happen.

\textit{Competing phases at $\nu=1$.---} We study interaction-induced competing phases at $\nu=1$ and show the quantum phase diagram in Fig.~\ref{fig:3}(a), which includes three distinct classes of states, a valley polarized (VP) state, spin density wave (SDW) states, and an in-plane ferromagnetic state (FM$_x$). We first describe the VP state, where the first moir\'e valence bands at $\pm K$ valleys have unequal occupations. Therefore, the VP state carries a finite out-of-plane spin polarization and spontaneously breaks $\mathcal{T}$ symmetry. Figure~\ref{fig:3}(f) shows a representative band structure calculated using the HF method for the VP state, which has a finite charge gap at $\nu=1$ and realizes a Chern insulator (ChI) with a quantized Chern number 1.  This gives rise to the VP-ChI phase in Fig.~\ref{fig:3}(a). The VP state may become topologically trivial for a large negative $V_{z\mathfrak{t}}$. However, such a VP {\it trivial} phase is absent in Fig.~\ref{fig:3}(a) because it is energetically unfavorable compared with SDW states for the parameter ranges used in this figure. To the other extreme ($V_{z\mathfrak{t}}\sim 0$), the VP state may become a metallic state that has partial valley polarization, though not energetically favorable compared with FM$_x$ in Fig.~\ref{fig:3}(a).

We now turn to SDW and FM$_x$ phases. The first moir\'e valence band in the topologically trivial regime of Fig.~\ref{fig:1}(d) can be described by a single-band tight-binding model on a triangular lattice, which, combined with the Coulomb interaction, realizes a generalized Hubbard model \cite{wu2018hubbard}. The single-band Hubbard model on a triangular lattice in the strong interaction limit has the $120^{\circ}$ antiferromagnetic N\'eel ground state \cite{wu2018hubbard,hu2021competing} at $\nu=1$, which spontaneously breaks the valley U(1) symmetry in our case and therefore, represents an intervalley-coherent state. The SDW in the $120^{\circ}$  N\'eel state spontaneously breaks the moir\'e translational symmetry, and has an expanded unit cell with a period $\sqrt{3} a_M$. In Figs.~\ref{fig:3}(b) -~\ref{fig:3}(d), we show three types of SDW states, distinguished by their charge and spin patterns in the bottom layer. In SDW1 and SDW2, the holes in the bottom layer are mainly concentrated on MM sites. Both SDW1 and SDW2 have the $120^{\circ}$  antiferromagnetic spin texture for holes on MM sites, but they differ by the spin vector chirality \cite{pan2020band}. In SDW3, the holes in the bottom layer are redistributed in moir\'e superlattices such that the hole density on MM and XX sites have comparable magnitude; the corresponding spin texture is shown in  Fig.~\ref{fig:3}(d). In addition, we also consider an FM$_x$ state (also an intervalley coherent state), where the spin for holes in the bottom is polarized along an in-plane direction. We perform a self-consistent HF calculation for the SDW and FM$_x$ states, compare their energies with the VP state, and obtain the phase diagram in Fig.~\ref{fig:3}(a) as a function of $V_{z\mathfrak{t}}$ and $\epsilon$.  

In Fig.~\ref{fig:3}(e), we show the charge gap as a function of $V_{z\mathfrak{t}}$ at a fixed $\epsilon$. As $V_{z\mathfrak{t}}$ increases from a large negative value toward 0, there is a first-order phase transition between the topologically trivial SDW phase and the topological VP ChI phase. Because the first-order phase transition breaks adiabatic continuity, the charge gap does not need to vanish at the transition point, and there can be a discontinuous jump in the gap. In contrast, the experimental charge gap \cite{li2021quantum} evolves {\it continuously} across the transition from a Mott insulator to a Chern insulator at $\nu=1$, which could be due to disorder effects~\cite{ahn2022disorderinduced}.

\begin{figure}[t]
    \centering
    \includegraphics[width=3.4in]{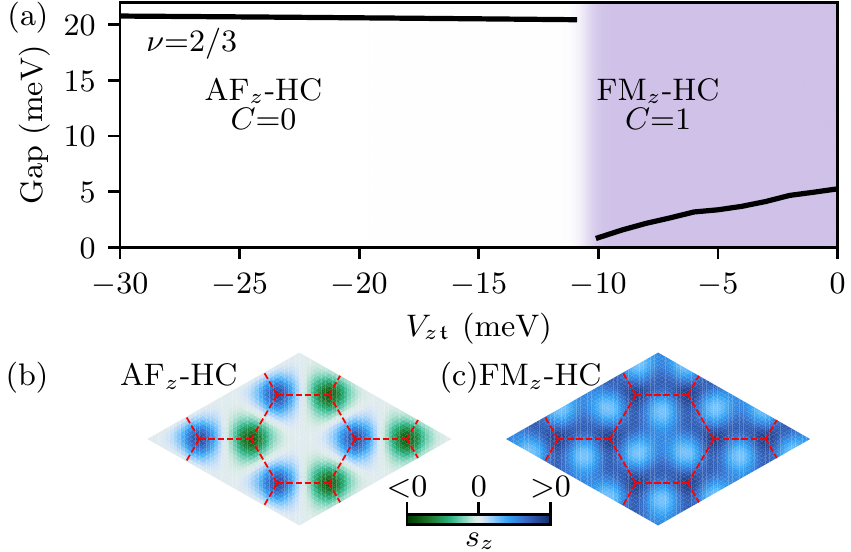}
    \caption{(a) Interaction-induced phase diagram at $\nu=2/3$ with $\epsilon=17$, $w$=12 meV, and $V_\mathfrak{b}$=7 meV. Note that the sample-to-gate distance $d$ has changed from 5 nm to 15 nm compared with other figures. The two competing phases are  AF$_z$-HC and FM$_z$-HC, which respectively have AF$_z$ and FM$_z$ spin ordering on a generalized Wigner crystal with honeycomb lattice on the bottom layer. (b), (c) The spatial variation of the $z$ component spin $s_z$ on the bottom layer for AF$_z$-HC and FM$_z$-HC. The red dashed lines mark the emergent honeycomb lattice with a period of $\sqrt{3} a_M$ and vertices on MM sites.}
    \label{fig:5}
\end{figure}

\textit{Topological charge density waves at $\nu=2/3$.---}
Correlated insulators can develop not only at integer $\nu$, but also at fractional $\nu$, as demonstrated experimentally in Refs.~\onlinecite{xu2020correlated,li2021imaging,liu2021excitonic,regan2020mott,zhou2021bilayer}.  Here we focus on $\nu=2/3$, where the Coulomb interaction can induce a charge density wave (CDW) state that forms a honeycomb lattice~\cite{pan2020quantum}. The enlarged unit cell has a period $\sqrt{3} a_M$ and contains three MM sites, but only two out of the three are occupied by holes in the bottom layer, which, therefore, leads to an effective honeycomb lattice. We theoretically find that the CDW state at $\nu=2/3$ is sensitive to the long-range part of the Coulomb interaction, and can be energetically stabilized for $d=15$ nm rather than $d=5$ nm, where $d$ is the sample-to-gate distance and controls the effective range of the Coulomb interaction. 

On the effective honeycomb lattice, there can be two types of spin ordering~\cite{pan2020quantum}: (1) antiferromagnetic (AF$_z$) order and (2) ferromagnetic (FM$_z$) order, where the spin polarization axis is out of the plane for both cases as illustrated respectively in Figs.~\ref{fig:5}(b) and~\ref{fig:5}(c). Here the FM$_z$ state is also valley polarized.  We perform self-consistent HF calculations respectively for the AF$_z$ honeycomb (AF$_z$-HC) state and the FM$_z$ honeycomb (FM$_z$-HC) state. Their energy competition gives rise to the $\nu=2/3$ phase diagram in Fig.~\ref{fig:5}(a), where the AF$_z$-HC and FM$_z$-HC states are favorable respectively for $V_{z\mathfrak{t}}$ below and above a critical value. Remarkably, when the FM$_z$-HC state becomes energetically favorable, it is also topologically nontrivial with a quantized Chern number 1. Therefore, there can be a topological CDW state at $\nu=2/3$.  Here the topology emerges with a mechanism similar to that of the VP ChI state at $\nu=1$, i.e., band inversion induced by $V_{z\mathfrak{t}}$. 

\textit{Conclusion.---} Our self-consistent theoretical results provide a reasonable qualitative understanding of the field-induced topological states in Ref.~\onlinecite{li2021quantum}. The theory sets up a general framework to study topological interaction effects in a transition metal dichalcogenide based moir\'e system, and can be readily applied to study a variety of interaction-induced symmetry breaking phases~\cite{wang2020correlated, xu2020correlated, gu2021dipolar,zhang2021correlated,pan2021interactiondriven,pan2020quantum,pan2020band,pan2022interaction}. For AB-stacked MoTe$_2$/WSe$_2$, many future directions can be explored in both theory and experiment. The $\nu=1$ Chern insulator has spontaneous VP, but does {\it not} break any {\it continuous} symmetry. It is an Ising-type ordering with a finite Curie temperature, which could be theoretically estimated by studying the valley magnon and domain fluctuations~\cite{wu2020collective}. This VP state can result in interesting optical phenomena, for example, Faraday and Kerr rotations~\cite{tse2010giant}. The multiple distinct quantum phases (SDW, ChI, and $\mathbb{Z}_2$ topological insulators) that emerge in a single system can be used as building blocks to design new quantum phases on demand. For example, by introducing proximitized superconductivity, an interface between the $\nu=1$ SDW state and the $\nu=2$ $\mathbb{Z}_2$ topological insulator could host Majorana zero modes. 

This work is supported by the National Key R$\&$D Program of China 2021YFA1401300 (F. W.) and the Laboratory for Physical Sciences (work at Maryland). We thank the University of Maryland High-Performance Computing Cluster (HPCC) for providing computational resources. F. W. also acknowledges support by startup funding from Wuhan University.

\bibliography{Paper_HeteroBilayer}

\begin{thebibliography}{74}%
\makeatletter
\providecommand \@ifxundefined [1]{%
 \@ifx{#1\undefined}
}%
\providecommand \@ifnum [1]{%
 \ifnum #1\expandafter \@firstoftwo
 \else \expandafter \@secondoftwo
 \fi
}%
\providecommand \@ifx [1]{%
 \ifx #1\expandafter \@firstoftwo
 \else \expandafter \@secondoftwo
 \fi
}%
\providecommand \natexlab [1]{#1}%
\providecommand \enquote  [1]{``#1''}%
\providecommand \bibnamefont  [1]{#1}%
\providecommand \bibfnamefont [1]{#1}%
\providecommand \citenamefont [1]{#1}%
\providecommand \href@noop [0]{\@secondoftwo}%
\providecommand \href [0]{\begingroup \@sanitize@url \@href}%
\providecommand \@href[1]{\@@startlink{#1}\@@href}%
\providecommand \@@href[1]{\endgroup#1\@@endlink}%
\providecommand \@sanitize@url [0]{\catcode `\\12\catcode `\$12\catcode
  `\&12\catcode `\#12\catcode `\^12\catcode `\_12\catcode `\%12\relax}%
\providecommand \@@startlink[1]{}%
\providecommand \@@endlink[0]{}%
\providecommand \url  [0]{\begingroup\@sanitize@url \@url }%
\providecommand \@url [1]{\endgroup\@href {#1}{\urlprefix }}%
\providecommand \urlprefix  [0]{URL }%
\providecommand \Eprint [0]{\href }%
\providecommand \doibase [0]{https://doi.org/}%
\providecommand \selectlanguage [0]{\@gobble}%
\providecommand \bibinfo  [0]{\@secondoftwo}%
\providecommand \bibfield  [0]{\@secondoftwo}%
\providecommand \translation [1]{[#1]}%
\providecommand \BibitemOpen [0]{}%
\providecommand \bibitemStop [0]{}%
\providecommand \bibitemNoStop [0]{.\EOS\space}%
\providecommand \EOS [0]{\spacefactor3000\relax}%
\providecommand \BibitemShut  [1]{\csname bibitem#1\endcsname}%
\let\auto@bib@innerbib\@empty
\bibitem [{\citenamefont {{von Klitzing}}\ \emph {et~al.}(1980)\citenamefont
  {{von Klitzing}}, \citenamefont {Dorda},\ and\ \citenamefont
  {Pepper}}]{vonklitzing1980new}%
  \BibitemOpen
  \bibfield  {author} {\bibinfo {author} {\bibfnamefont {K.}~\bibnamefont {{von
  Klitzing}}}, \bibinfo {author} {\bibfnamefont {G.}~\bibnamefont {Dorda}},\
  and\ \bibinfo {author} {\bibfnamefont {M.}~\bibnamefont {Pepper}},\
  }\bibfield  {title} {\bibinfo {title} {New {{Method}} for {{High-Accuracy
  Determination}} of the {{Fine-Structure Constant Based}} on {{Quantized Hall
  Resistance}}},\ }\href {https://doi.org/10.1103/PhysRevLett.45.494}
  {\bibfield  {journal} {\bibinfo  {journal} {Phys. Rev. Lett.}\ }\textbf
  {\bibinfo {volume} {45}},\ \bibinfo {pages} {494} (\bibinfo {year}
  {1980})}\BibitemShut {NoStop}%
\bibitem [{\citenamefont {Thouless}\ \emph {et~al.}(1982)\citenamefont
  {Thouless}, \citenamefont {Kohmoto}, \citenamefont {Nightingale},\ and\
  \citenamefont {{den Nijs}}}]{thouless1982quantized}%
  \BibitemOpen
  \bibfield  {author} {\bibinfo {author} {\bibfnamefont {D.~J.}\ \bibnamefont
  {Thouless}}, \bibinfo {author} {\bibfnamefont {M.}~\bibnamefont {Kohmoto}},
  \bibinfo {author} {\bibfnamefont {M.~P.}\ \bibnamefont {Nightingale}},\ and\
  \bibinfo {author} {\bibfnamefont {M.}~\bibnamefont {{den Nijs}}},\ }\bibfield
   {title} {\bibinfo {title} {Quantized {{Hall Conductance}} in a
  {{Two-Dimensional Periodic Potential}}},\ }\href
  {https://doi.org/10.1103/PhysRevLett.49.405} {\bibfield  {journal} {\bibinfo
  {journal} {Phys. Rev. Lett.}\ }\textbf {\bibinfo {volume} {49}},\ \bibinfo
  {pages} {405} (\bibinfo {year} {1982})}\BibitemShut {NoStop}%
\bibitem [{\citenamefont {Niu}\ \emph {et~al.}(1985)\citenamefont {Niu},
  \citenamefont {Thouless},\ and\ \citenamefont {Wu}}]{niu1985quantized}%
  \BibitemOpen
  \bibfield  {author} {\bibinfo {author} {\bibfnamefont {Q.}~\bibnamefont
  {Niu}}, \bibinfo {author} {\bibfnamefont {D.~J.}\ \bibnamefont {Thouless}},\
  and\ \bibinfo {author} {\bibfnamefont {Y.-S.}\ \bibnamefont {Wu}},\
  }\bibfield  {title} {\bibinfo {title} {Quantized {{Hall}} conductance as a
  topological invariant},\ }\href {https://doi.org/10.1103/PhysRevB.31.3372}
  {\bibfield  {journal} {\bibinfo  {journal} {Phys. Rev. B}\ }\textbf {\bibinfo
  {volume} {31}},\ \bibinfo {pages} {3372} (\bibinfo {year}
  {1985})}\BibitemShut {NoStop}%
\bibitem [{\citenamefont {Haldane}(1988)}]{haldane1988model}%
  \BibitemOpen
  \bibfield  {author} {\bibinfo {author} {\bibfnamefont {F.~D.~M.}\
  \bibnamefont {Haldane}},\ }\bibfield  {title} {\bibinfo {title} {Model for a
  {{Quantum Hall Effect}} without {{Landau Levels}}: {{Condensed-Matter
  Realization}} of the "{{Parity Anomaly}}"},\ }\href
  {https://doi.org/10.1103/PhysRevLett.61.2015} {\bibfield  {journal} {\bibinfo
   {journal} {Phys. Rev. Lett.}\ }\textbf {\bibinfo {volume} {61}},\ \bibinfo
  {pages} {2015} (\bibinfo {year} {1988})}\BibitemShut {NoStop}%
\bibitem [{\citenamefont {Kane}\ and\ \citenamefont
  {Mele}(2005)}]{kane2005quantum}%
  \BibitemOpen
  \bibfield  {author} {\bibinfo {author} {\bibfnamefont {C.~L.}\ \bibnamefont
  {Kane}}\ and\ \bibinfo {author} {\bibfnamefont {E.~J.}\ \bibnamefont
  {Mele}},\ }\bibfield  {title} {\bibinfo {title} {Quantum {{Spin Hall Effect}}
  in {{Graphene}}},\ }\href {https://doi.org/10.1103/PhysRevLett.95.226801}
  {\bibfield  {journal} {\bibinfo  {journal} {Phys. Rev. Lett.}\ }\textbf
  {\bibinfo {volume} {95}},\ \bibinfo {pages} {226801} (\bibinfo {year}
  {2005})}\BibitemShut {NoStop}%
\bibitem [{\citenamefont {Onoda}\ and\ \citenamefont
  {Nagaosa}(2003)}]{onoda2003quantized}%
  \BibitemOpen
  \bibfield  {author} {\bibinfo {author} {\bibfnamefont {M.}~\bibnamefont
  {Onoda}}\ and\ \bibinfo {author} {\bibfnamefont {N.}~\bibnamefont
  {Nagaosa}},\ }\bibfield  {title} {\bibinfo {title} {Quantized {{Anomalous
  Hall Effect}} in {{Two-Dimensional Ferromagnets}}: {{Quantum Hall Effect}} in
  {{Metals}}},\ }\href {https://doi.org/10.1103/PhysRevLett.90.206601}
  {\bibfield  {journal} {\bibinfo  {journal} {Phys. Rev. Lett.}\ }\textbf
  {\bibinfo {volume} {90}},\ \bibinfo {pages} {206601} (\bibinfo {year}
  {2003})}\BibitemShut {NoStop}%
\bibitem [{\citenamefont {Qi}\ \emph {et~al.}(2006)\citenamefont {Qi},
  \citenamefont {Wu},\ and\ \citenamefont {Zhang}}]{qi2006topological}%
  \BibitemOpen
  \bibfield  {author} {\bibinfo {author} {\bibfnamefont {X.-L.}\ \bibnamefont
  {Qi}}, \bibinfo {author} {\bibfnamefont {Y.-S.}\ \bibnamefont {Wu}},\ and\
  \bibinfo {author} {\bibfnamefont {S.-C.}\ \bibnamefont {Zhang}},\ }\bibfield
  {title} {\bibinfo {title} {Topological quantization of the spin {{Hall}}
  effect in two-dimensional paramagnetic semiconductors},\ }\href
  {https://doi.org/10.1103/PhysRevB.74.085308} {\bibfield  {journal} {\bibinfo
  {journal} {Phys. Rev. B}\ }\textbf {\bibinfo {volume} {74}},\ \bibinfo
  {pages} {085308} (\bibinfo {year} {2006})}\BibitemShut {NoStop}%
\bibitem [{\citenamefont {Qi}\ \emph {et~al.}(2008)\citenamefont {Qi},
  \citenamefont {Hughes},\ and\ \citenamefont {Zhang}}]{qi2008topological}%
  \BibitemOpen
  \bibfield  {author} {\bibinfo {author} {\bibfnamefont {X.-L.}\ \bibnamefont
  {Qi}}, \bibinfo {author} {\bibfnamefont {T.~L.}\ \bibnamefont {Hughes}},\
  and\ \bibinfo {author} {\bibfnamefont {S.-C.}\ \bibnamefont {Zhang}},\
  }\bibfield  {title} {\bibinfo {title} {Topological field theory of
  time-reversal invariant insulators},\ }\href
  {https://doi.org/10.1103/PhysRevB.78.195424} {\bibfield  {journal} {\bibinfo
  {journal} {Phys. Rev. B}\ }\textbf {\bibinfo {volume} {78}},\ \bibinfo
  {pages} {195424} (\bibinfo {year} {2008})}\BibitemShut {NoStop}%
\bibitem [{\citenamefont {Liu}\ \emph {et~al.}(2008{\natexlab{a}})\citenamefont
  {Liu}, \citenamefont {Qi}, \citenamefont {Dai}, \citenamefont {Fang},\ and\
  \citenamefont {Zhang}}]{liu2008quantum}%
  \BibitemOpen
  \bibfield  {author} {\bibinfo {author} {\bibfnamefont {C.-X.}\ \bibnamefont
  {Liu}}, \bibinfo {author} {\bibfnamefont {X.-L.}\ \bibnamefont {Qi}},
  \bibinfo {author} {\bibfnamefont {X.}~\bibnamefont {Dai}}, \bibinfo {author}
  {\bibfnamefont {Z.}~\bibnamefont {Fang}},\ and\ \bibinfo {author}
  {\bibfnamefont {S.-C.}\ \bibnamefont {Zhang}},\ }\bibfield  {title} {\bibinfo
  {title} {Quantum {{Anomalous Hall Effect}} in
  {${\mathrm{Hg}}_{1\ensuremath{-}y}{\mathrm{Mn}}_{y}\mathrm{Te}$} {{Quantum
  Wells}}},\ }\href {https://doi.org/10.1103/PhysRevLett.101.146802} {\bibfield
   {journal} {\bibinfo  {journal} {Phys. Rev. Lett.}\ }\textbf {\bibinfo
  {volume} {101}},\ \bibinfo {pages} {146802} (\bibinfo {year}
  {2008}{\natexlab{a}})}\BibitemShut {NoStop}%
\bibitem [{\citenamefont {Yu}\ \emph {et~al.}(2010)\citenamefont {Yu},
  \citenamefont {Zhang}, \citenamefont {Zhang}, \citenamefont {Zhang},
  \citenamefont {Dai},\ and\ \citenamefont {Fang}}]{yu2010quantized}%
  \BibitemOpen
  \bibfield  {author} {\bibinfo {author} {\bibfnamefont {R.}~\bibnamefont
  {Yu}}, \bibinfo {author} {\bibfnamefont {W.}~\bibnamefont {Zhang}}, \bibinfo
  {author} {\bibfnamefont {H.-J.}\ \bibnamefont {Zhang}}, \bibinfo {author}
  {\bibfnamefont {S.-C.}\ \bibnamefont {Zhang}}, \bibinfo {author}
  {\bibfnamefont {X.}~\bibnamefont {Dai}},\ and\ \bibinfo {author}
  {\bibfnamefont {Z.}~\bibnamefont {Fang}},\ }\bibfield  {title} {\bibinfo
  {title} {Quantized {{Anomalous Hall Effect}} in {{Magnetic Topological
  Insulators}}},\ }\href {https://doi.org/10.1126/science.1187485} {\bibfield
  {journal} {\bibinfo  {journal} {Science}\ }\textbf {\bibinfo {volume}
  {329}},\ \bibinfo {pages} {61} (\bibinfo {year} {2010})}\BibitemShut
  {NoStop}%
\bibitem [{\citenamefont {Qiao}\ \emph {et~al.}(2010)\citenamefont {Qiao},
  \citenamefont {Yang}, \citenamefont {Feng}, \citenamefont {Tse},
  \citenamefont {Ding}, \citenamefont {Yao}, \citenamefont {Wang},\ and\
  \citenamefont {Niu}}]{qiao2010quantum}%
  \BibitemOpen
  \bibfield  {author} {\bibinfo {author} {\bibfnamefont {Z.}~\bibnamefont
  {Qiao}}, \bibinfo {author} {\bibfnamefont {S.~A.}\ \bibnamefont {Yang}},
  \bibinfo {author} {\bibfnamefont {W.}~\bibnamefont {Feng}}, \bibinfo {author}
  {\bibfnamefont {W.-K.}\ \bibnamefont {Tse}}, \bibinfo {author} {\bibfnamefont
  {J.}~\bibnamefont {Ding}}, \bibinfo {author} {\bibfnamefont {Y.}~\bibnamefont
  {Yao}}, \bibinfo {author} {\bibfnamefont {J.}~\bibnamefont {Wang}},\ and\
  \bibinfo {author} {\bibfnamefont {Q.}~\bibnamefont {Niu}},\ }\bibfield
  {title} {\bibinfo {title} {Quantum anomalous {{Hall}} effect in graphene from
  {{Rashba}} and exchange effects},\ }\href
  {https://doi.org/10.1103/PhysRevB.82.161414} {\bibfield  {journal} {\bibinfo
  {journal} {Phys. Rev. B}\ }\textbf {\bibinfo {volume} {82}},\ \bibinfo
  {pages} {161414} (\bibinfo {year} {2010})}\BibitemShut {NoStop}%
\bibitem [{\citenamefont {Nomura}\ and\ \citenamefont
  {Nagaosa}(2011)}]{nomura2011surfacequantized}%
  \BibitemOpen
  \bibfield  {author} {\bibinfo {author} {\bibfnamefont {K.}~\bibnamefont
  {Nomura}}\ and\ \bibinfo {author} {\bibfnamefont {N.}~\bibnamefont
  {Nagaosa}},\ }\bibfield  {title} {\bibinfo {title} {Surface-{{Quantized
  Anomalous Hall Current}} and the {{Magnetoelectric Effect}} in {{Magnetically
  Disordered Topological Insulators}}},\ }\href
  {https://doi.org/10.1103/PhysRevLett.106.166802} {\bibfield  {journal}
  {\bibinfo  {journal} {Phys. Rev. Lett.}\ }\textbf {\bibinfo {volume} {106}},\
  \bibinfo {pages} {166802} (\bibinfo {year} {2011})}\BibitemShut {NoStop}%
\bibitem [{\citenamefont {Zhang}\ \emph
  {et~al.}(2019{\natexlab{a}})\citenamefont {Zhang}, \citenamefont {Shi},
  \citenamefont {Zhu}, \citenamefont {Xing}, \citenamefont {Zhang},\ and\
  \citenamefont {Wang}}]{zhang2019topological}%
  \BibitemOpen
  \bibfield  {author} {\bibinfo {author} {\bibfnamefont {D.}~\bibnamefont
  {Zhang}}, \bibinfo {author} {\bibfnamefont {M.}~\bibnamefont {Shi}}, \bibinfo
  {author} {\bibfnamefont {T.}~\bibnamefont {Zhu}}, \bibinfo {author}
  {\bibfnamefont {D.}~\bibnamefont {Xing}}, \bibinfo {author} {\bibfnamefont
  {H.}~\bibnamefont {Zhang}},\ and\ \bibinfo {author} {\bibfnamefont
  {J.}~\bibnamefont {Wang}},\ }\bibfield  {title} {\bibinfo {title}
  {Topological {{Axion States}} in the {{Magnetic Insulator}}
  {${\mathrm{MnBi}}_{2}{\mathrm{Te}}_{4}$} with the {{Quantized Magnetoelectric
  Effect}}},\ }\href {https://doi.org/10.1103/PhysRevLett.122.206401}
  {\bibfield  {journal} {\bibinfo  {journal} {Phys. Rev. Lett.}\ }\textbf
  {\bibinfo {volume} {122}},\ \bibinfo {pages} {206401} (\bibinfo {year}
  {2019}{\natexlab{a}})}\BibitemShut {NoStop}%
\bibitem [{\citenamefont {Li}\ \emph {et~al.}(2019)\citenamefont {Li},
  \citenamefont {Li}, \citenamefont {Du}, \citenamefont {Wang}, \citenamefont
  {Gu}, \citenamefont {Zhang}, \citenamefont {He}, \citenamefont {Duan},\ and\
  \citenamefont {Xu}}]{li2019intrinsic}%
  \BibitemOpen
  \bibfield  {author} {\bibinfo {author} {\bibfnamefont {J.}~\bibnamefont
  {Li}}, \bibinfo {author} {\bibfnamefont {Y.}~\bibnamefont {Li}}, \bibinfo
  {author} {\bibfnamefont {S.}~\bibnamefont {Du}}, \bibinfo {author}
  {\bibfnamefont {Z.}~\bibnamefont {Wang}}, \bibinfo {author} {\bibfnamefont
  {B.-L.}\ \bibnamefont {Gu}}, \bibinfo {author} {\bibfnamefont {S.-C.}\
  \bibnamefont {Zhang}}, \bibinfo {author} {\bibfnamefont {K.}~\bibnamefont
  {He}}, \bibinfo {author} {\bibfnamefont {W.}~\bibnamefont {Duan}},\ and\
  \bibinfo {author} {\bibfnamefont {Y.}~\bibnamefont {Xu}},\ }\bibfield
  {title} {\bibinfo {title} {Intrinsic magnetic topological insulators in van
  der {{Waals}} layered {{MnBi2Te4-family}} materials},\ }\href
  {https://doi.org/10.1126/sciadv.aaw5685} {\bibfield  {journal} {\bibinfo
  {journal} {Science Advances}\ }\textbf {\bibinfo {volume} {5}},\ \bibinfo
  {pages} {eaaw5685} (\bibinfo {year} {2019})}\BibitemShut {NoStop}%
\bibitem [{\citenamefont {Otrokov}\ \emph {et~al.}(2019)\citenamefont
  {Otrokov}, \citenamefont {Klimovskikh}, \citenamefont {Bentmann},
  \citenamefont {Estyunin}, \citenamefont {Zeugner}, \citenamefont {Aliev},
  \citenamefont {Ga{\ss}}, \citenamefont {Wolter}, \citenamefont {Koroleva},
  \citenamefont {Shikin}, \citenamefont {{Blanco-Rey}}, \citenamefont
  {Hoffmann}, \citenamefont {Rusinov}, \citenamefont {Vyazovskaya},
  \citenamefont {Eremeev}, \citenamefont {Koroteev}, \citenamefont {Kuznetsov},
  \citenamefont {Freyse}, \citenamefont {{S{\'a}nchez-Barriga}}, \citenamefont
  {Amiraslanov}, \citenamefont {Babanly}, \citenamefont {Mamedov},
  \citenamefont {Abdullayev}, \citenamefont {Zverev}, \citenamefont {Alfonsov},
  \citenamefont {Kataev}, \citenamefont {B{\"u}chner}, \citenamefont {Schwier},
  \citenamefont {Kumar}, \citenamefont {Kimura}, \citenamefont {Petaccia},
  \citenamefont {Di~Santo}, \citenamefont {Vidal}, \citenamefont {Schatz},
  \citenamefont {Ki{\ss}ner}, \citenamefont {{\"U}nzelmann}, \citenamefont
  {Min}, \citenamefont {Moser}, \citenamefont {Peixoto}, \citenamefont
  {Reinert}, \citenamefont {Ernst}, \citenamefont {Echenique}, \citenamefont
  {Isaeva},\ and\ \citenamefont {Chulkov}}]{otrokov2019prediction}%
  \BibitemOpen
  \bibfield  {author} {\bibinfo {author} {\bibfnamefont {M.~M.}\ \bibnamefont
  {Otrokov}}, \bibinfo {author} {\bibfnamefont {I.~I.}\ \bibnamefont
  {Klimovskikh}}, \bibinfo {author} {\bibfnamefont {H.}~\bibnamefont
  {Bentmann}}, \bibinfo {author} {\bibfnamefont {D.}~\bibnamefont {Estyunin}},
  \bibinfo {author} {\bibfnamefont {A.}~\bibnamefont {Zeugner}}, \bibinfo
  {author} {\bibfnamefont {Z.~S.}\ \bibnamefont {Aliev}}, \bibinfo {author}
  {\bibfnamefont {S.}~\bibnamefont {Ga{\ss}}}, \bibinfo {author} {\bibfnamefont
  {A.~U.~B.}\ \bibnamefont {Wolter}}, \bibinfo {author} {\bibfnamefont {A.~V.}\
  \bibnamefont {Koroleva}}, \bibinfo {author} {\bibfnamefont {A.~M.}\
  \bibnamefont {Shikin}}, \bibinfo {author} {\bibfnamefont {M.}~\bibnamefont
  {{Blanco-Rey}}}, \bibinfo {author} {\bibfnamefont {M.}~\bibnamefont
  {Hoffmann}}, \bibinfo {author} {\bibfnamefont {I.~P.}\ \bibnamefont
  {Rusinov}}, \bibinfo {author} {\bibfnamefont {A.~Y.}\ \bibnamefont
  {Vyazovskaya}}, \bibinfo {author} {\bibfnamefont {S.~V.}\ \bibnamefont
  {Eremeev}}, \bibinfo {author} {\bibfnamefont {Y.~M.}\ \bibnamefont
  {Koroteev}}, \bibinfo {author} {\bibfnamefont {V.~M.}\ \bibnamefont
  {Kuznetsov}}, \bibinfo {author} {\bibfnamefont {F.}~\bibnamefont {Freyse}},
  \bibinfo {author} {\bibfnamefont {J.}~\bibnamefont {{S{\'a}nchez-Barriga}}},
  \bibinfo {author} {\bibfnamefont {I.~R.}\ \bibnamefont {Amiraslanov}},
  \bibinfo {author} {\bibfnamefont {M.~B.}\ \bibnamefont {Babanly}}, \bibinfo
  {author} {\bibfnamefont {N.~T.}\ \bibnamefont {Mamedov}}, \bibinfo {author}
  {\bibfnamefont {N.~A.}\ \bibnamefont {Abdullayev}}, \bibinfo {author}
  {\bibfnamefont {V.~N.}\ \bibnamefont {Zverev}}, \bibinfo {author}
  {\bibfnamefont {A.}~\bibnamefont {Alfonsov}}, \bibinfo {author}
  {\bibfnamefont {V.}~\bibnamefont {Kataev}}, \bibinfo {author} {\bibfnamefont
  {B.}~\bibnamefont {B{\"u}chner}}, \bibinfo {author} {\bibfnamefont {E.~F.}\
  \bibnamefont {Schwier}}, \bibinfo {author} {\bibfnamefont {S.}~\bibnamefont
  {Kumar}}, \bibinfo {author} {\bibfnamefont {A.}~\bibnamefont {Kimura}},
  \bibinfo {author} {\bibfnamefont {L.}~\bibnamefont {Petaccia}}, \bibinfo
  {author} {\bibfnamefont {G.}~\bibnamefont {Di~Santo}}, \bibinfo {author}
  {\bibfnamefont {R.~C.}\ \bibnamefont {Vidal}}, \bibinfo {author}
  {\bibfnamefont {S.}~\bibnamefont {Schatz}}, \bibinfo {author} {\bibfnamefont
  {K.}~\bibnamefont {Ki{\ss}ner}}, \bibinfo {author} {\bibfnamefont
  {M.}~\bibnamefont {{\"U}nzelmann}}, \bibinfo {author} {\bibfnamefont {C.~H.}\
  \bibnamefont {Min}}, \bibinfo {author} {\bibfnamefont {S.}~\bibnamefont
  {Moser}}, \bibinfo {author} {\bibfnamefont {T.~R.~F.}\ \bibnamefont
  {Peixoto}}, \bibinfo {author} {\bibfnamefont {F.}~\bibnamefont {Reinert}},
  \bibinfo {author} {\bibfnamefont {A.}~\bibnamefont {Ernst}}, \bibinfo
  {author} {\bibfnamefont {P.~M.}\ \bibnamefont {Echenique}}, \bibinfo {author}
  {\bibfnamefont {A.}~\bibnamefont {Isaeva}},\ and\ \bibinfo {author}
  {\bibfnamefont {E.~V.}\ \bibnamefont {Chulkov}},\ }\bibfield  {title}
  {\bibinfo {title} {Prediction and observation of an antiferromagnetic
  topological insulator},\ }\href {https://doi.org/10.1038/s41586-019-1840-9}
  {\bibfield  {journal} {\bibinfo  {journal} {Nature}\ }\textbf {\bibinfo
  {volume} {576}},\ \bibinfo {pages} {416} (\bibinfo {year}
  {2019})}\BibitemShut {NoStop}%
\bibitem [{\citenamefont {Liu}\ \emph {et~al.}(2008{\natexlab{b}})\citenamefont
  {Liu}, \citenamefont {Hughes}, \citenamefont {Qi}, \citenamefont {Wang},\
  and\ \citenamefont {Zhang}}]{liu2008quantuma}%
  \BibitemOpen
  \bibfield  {author} {\bibinfo {author} {\bibfnamefont {C.}~\bibnamefont
  {Liu}}, \bibinfo {author} {\bibfnamefont {T.~L.}\ \bibnamefont {Hughes}},
  \bibinfo {author} {\bibfnamefont {X.-L.}\ \bibnamefont {Qi}}, \bibinfo
  {author} {\bibfnamefont {K.}~\bibnamefont {Wang}},\ and\ \bibinfo {author}
  {\bibfnamefont {S.-C.}\ \bibnamefont {Zhang}},\ }\bibfield  {title} {\bibinfo
  {title} {Quantum {{Spin Hall Effect}} in {{Inverted Type-II
  Semiconductors}}},\ }\href {https://doi.org/10.1103/PhysRevLett.100.236601}
  {\bibfield  {journal} {\bibinfo  {journal} {Phys. Rev. Lett.}\ }\textbf
  {\bibinfo {volume} {100}},\ \bibinfo {pages} {236601} (\bibinfo {year}
  {2008}{\natexlab{b}})}\BibitemShut {NoStop}%
\bibitem [{\citenamefont {Zhang}\ \emph
  {et~al.}(2019{\natexlab{b}})\citenamefont {Zhang}, \citenamefont {Mao},
  \citenamefont {Cao}, \citenamefont {{Jarillo-Herrero}},\ and\ \citenamefont
  {Senthil}}]{zhang2019nearly}%
  \BibitemOpen
  \bibfield  {author} {\bibinfo {author} {\bibfnamefont {Y.-H.}\ \bibnamefont
  {Zhang}}, \bibinfo {author} {\bibfnamefont {D.}~\bibnamefont {Mao}}, \bibinfo
  {author} {\bibfnamefont {Y.}~\bibnamefont {Cao}}, \bibinfo {author}
  {\bibfnamefont {P.}~\bibnamefont {{Jarillo-Herrero}}},\ and\ \bibinfo
  {author} {\bibfnamefont {T.}~\bibnamefont {Senthil}},\ }\bibfield  {title}
  {\bibinfo {title} {Nearly flat {{Chern}} bands in moir\'e superlattices},\
  }\href {https://doi.org/10.1103/PhysRevB.99.075127} {\bibfield  {journal}
  {\bibinfo  {journal} {Phys. Rev. B}\ }\textbf {\bibinfo {volume} {99}},\
  \bibinfo {pages} {075127} (\bibinfo {year} {2019}{\natexlab{b}})}\BibitemShut
  {NoStop}%
\bibitem [{\citenamefont {Bernevig}\ and\ \citenamefont
  {Zhang}(2006)}]{bernevig2006quantum}%
  \BibitemOpen
  \bibfield  {author} {\bibinfo {author} {\bibfnamefont {B.~A.}\ \bibnamefont
  {Bernevig}}\ and\ \bibinfo {author} {\bibfnamefont {S.-C.}\ \bibnamefont
  {Zhang}},\ }\bibfield  {title} {\bibinfo {title} {Quantum {{Spin Hall
  Effect}}},\ }\href {https://doi.org/10.1103/PhysRevLett.96.106802} {\bibfield
   {journal} {\bibinfo  {journal} {Phys. Rev. Lett.}\ }\textbf {\bibinfo
  {volume} {96}},\ \bibinfo {pages} {106802} (\bibinfo {year}
  {2006})}\BibitemShut {NoStop}%
\bibitem [{\citenamefont {Sheng}\ \emph {et~al.}(2006)\citenamefont {Sheng},
  \citenamefont {Weng}, \citenamefont {Sheng},\ and\ \citenamefont
  {Haldane}}]{sheng2006quantum}%
  \BibitemOpen
  \bibfield  {author} {\bibinfo {author} {\bibfnamefont {D.~N.}\ \bibnamefont
  {Sheng}}, \bibinfo {author} {\bibfnamefont {Z.~Y.}\ \bibnamefont {Weng}},
  \bibinfo {author} {\bibfnamefont {L.}~\bibnamefont {Sheng}},\ and\ \bibinfo
  {author} {\bibfnamefont {F.~D.~M.}\ \bibnamefont {Haldane}},\ }\bibfield
  {title} {\bibinfo {title} {Quantum {{Spin-Hall Effect}} and {{Topologically
  Invariant Chern Numbers}}},\ }\href
  {https://doi.org/10.1103/PhysRevLett.97.036808} {\bibfield  {journal}
  {\bibinfo  {journal} {Phys. Rev. Lett.}\ }\textbf {\bibinfo {volume} {97}},\
  \bibinfo {pages} {036808} (\bibinfo {year} {2006})}\BibitemShut {NoStop}%
\bibitem [{\citenamefont {Qian}\ \emph {et~al.}(2014)\citenamefont {Qian},
  \citenamefont {Liu}, \citenamefont {Fu},\ and\ \citenamefont
  {Li}}]{qian2014quantum}%
  \BibitemOpen
  \bibfield  {author} {\bibinfo {author} {\bibfnamefont {X.}~\bibnamefont
  {Qian}}, \bibinfo {author} {\bibfnamefont {J.}~\bibnamefont {Liu}}, \bibinfo
  {author} {\bibfnamefont {L.}~\bibnamefont {Fu}},\ and\ \bibinfo {author}
  {\bibfnamefont {J.}~\bibnamefont {Li}},\ }\bibfield  {title} {\bibinfo
  {title} {Quantum spin {{Hall}} effect in two-dimensional transition metal
  dichalcogenides},\ }\href {https://doi.org/10.1126/science.1256815}
  {\bibfield  {journal} {\bibinfo  {journal} {Science}\ }\textbf {\bibinfo
  {volume} {346}},\ \bibinfo {pages} {1344} (\bibinfo {year}
  {2014})}\BibitemShut {NoStop}%
\bibitem [{\citenamefont {Chang}\ \emph {et~al.}(2013)\citenamefont {Chang},
  \citenamefont {Zhang}, \citenamefont {Feng}, \citenamefont {Shen},
  \citenamefont {Zhang}, \citenamefont {Guo}, \citenamefont {Li}, \citenamefont
  {Ou}, \citenamefont {Wei}, \citenamefont {Wang}, \citenamefont {Ji},
  \citenamefont {Feng}, \citenamefont {Ji}, \citenamefont {Chen}, \citenamefont
  {Jia}, \citenamefont {Dai}, \citenamefont {Fang}, \citenamefont {Zhang},
  \citenamefont {He}, \citenamefont {Wang}, \citenamefont {Lu}, \citenamefont
  {Ma},\ and\ \citenamefont {Xue}}]{chang2013experimental}%
  \BibitemOpen
  \bibfield  {author} {\bibinfo {author} {\bibfnamefont {C.-Z.}\ \bibnamefont
  {Chang}}, \bibinfo {author} {\bibfnamefont {J.}~\bibnamefont {Zhang}},
  \bibinfo {author} {\bibfnamefont {X.}~\bibnamefont {Feng}}, \bibinfo {author}
  {\bibfnamefont {J.}~\bibnamefont {Shen}}, \bibinfo {author} {\bibfnamefont
  {Z.}~\bibnamefont {Zhang}}, \bibinfo {author} {\bibfnamefont
  {M.}~\bibnamefont {Guo}}, \bibinfo {author} {\bibfnamefont {K.}~\bibnamefont
  {Li}}, \bibinfo {author} {\bibfnamefont {Y.}~\bibnamefont {Ou}}, \bibinfo
  {author} {\bibfnamefont {P.}~\bibnamefont {Wei}}, \bibinfo {author}
  {\bibfnamefont {L.-L.}\ \bibnamefont {Wang}}, \bibinfo {author}
  {\bibfnamefont {Z.-Q.}\ \bibnamefont {Ji}}, \bibinfo {author} {\bibfnamefont
  {Y.}~\bibnamefont {Feng}}, \bibinfo {author} {\bibfnamefont {S.}~\bibnamefont
  {Ji}}, \bibinfo {author} {\bibfnamefont {X.}~\bibnamefont {Chen}}, \bibinfo
  {author} {\bibfnamefont {J.}~\bibnamefont {Jia}}, \bibinfo {author}
  {\bibfnamefont {X.}~\bibnamefont {Dai}}, \bibinfo {author} {\bibfnamefont
  {Z.}~\bibnamefont {Fang}}, \bibinfo {author} {\bibfnamefont {S.-C.}\
  \bibnamefont {Zhang}}, \bibinfo {author} {\bibfnamefont {K.}~\bibnamefont
  {He}}, \bibinfo {author} {\bibfnamefont {Y.}~\bibnamefont {Wang}}, \bibinfo
  {author} {\bibfnamefont {L.}~\bibnamefont {Lu}}, \bibinfo {author}
  {\bibfnamefont {X.-C.}\ \bibnamefont {Ma}},\ and\ \bibinfo {author}
  {\bibfnamefont {Q.-K.}\ \bibnamefont {Xue}},\ }\bibfield  {title} {\bibinfo
  {title} {Experimental {{Observation}} of the {{Quantum Anomalous Hall
  Effect}} in a {{Magnetic Topological Insulator}}},\ }\href
  {https://doi.org/10.1126/science.1234414} {\bibfield  {journal} {\bibinfo
  {journal} {Science}\ }\textbf {\bibinfo {volume} {340}},\ \bibinfo {pages}
  {167} (\bibinfo {year} {2013})}\BibitemShut {NoStop}%
\bibitem [{\citenamefont {Zhao}\ \emph {et~al.}(2020)\citenamefont {Zhao},
  \citenamefont {Zhang}, \citenamefont {Mei}, \citenamefont {Zhou},
  \citenamefont {Yi}, \citenamefont {Zhang}, \citenamefont {Yu}, \citenamefont
  {Xiao}, \citenamefont {Wang}, \citenamefont {Samarth}, \citenamefont {Chan},
  \citenamefont {Liu},\ and\ \citenamefont {Chang}}]{zhao2020tuning}%
  \BibitemOpen
  \bibfield  {author} {\bibinfo {author} {\bibfnamefont {Y.-F.}\ \bibnamefont
  {Zhao}}, \bibinfo {author} {\bibfnamefont {R.}~\bibnamefont {Zhang}},
  \bibinfo {author} {\bibfnamefont {R.}~\bibnamefont {Mei}}, \bibinfo {author}
  {\bibfnamefont {L.-J.}\ \bibnamefont {Zhou}}, \bibinfo {author}
  {\bibfnamefont {H.}~\bibnamefont {Yi}}, \bibinfo {author} {\bibfnamefont
  {Y.-Q.}\ \bibnamefont {Zhang}}, \bibinfo {author} {\bibfnamefont
  {J.}~\bibnamefont {Yu}}, \bibinfo {author} {\bibfnamefont {R.}~\bibnamefont
  {Xiao}}, \bibinfo {author} {\bibfnamefont {K.}~\bibnamefont {Wang}}, \bibinfo
  {author} {\bibfnamefont {N.}~\bibnamefont {Samarth}}, \bibinfo {author}
  {\bibfnamefont {M.~H.~W.}\ \bibnamefont {Chan}}, \bibinfo {author}
  {\bibfnamefont {C.-X.}\ \bibnamefont {Liu}},\ and\ \bibinfo {author}
  {\bibfnamefont {C.-Z.}\ \bibnamefont {Chang}},\ }\bibfield  {title} {\bibinfo
  {title} {Tuning the {{Chern}} number in quantum anomalous {{Hall}}
  insulators},\ }\href {https://doi.org/10.1038/s41586-020-3020-3} {\bibfield
  {journal} {\bibinfo  {journal} {Nature}\ }\textbf {\bibinfo {volume} {588}},\
  \bibinfo {pages} {419} (\bibinfo {year} {2020})}\BibitemShut {NoStop}%
\bibitem [{\citenamefont {Deng}\ \emph {et~al.}(2020)\citenamefont {Deng},
  \citenamefont {Yu}, \citenamefont {Shi}, \citenamefont {Guo}, \citenamefont
  {Xu}, \citenamefont {Wang}, \citenamefont {Chen},\ and\ \citenamefont
  {Zhang}}]{deng2020quantum}%
  \BibitemOpen
  \bibfield  {author} {\bibinfo {author} {\bibfnamefont {Y.}~\bibnamefont
  {Deng}}, \bibinfo {author} {\bibfnamefont {Y.}~\bibnamefont {Yu}}, \bibinfo
  {author} {\bibfnamefont {M.~Z.}\ \bibnamefont {Shi}}, \bibinfo {author}
  {\bibfnamefont {Z.}~\bibnamefont {Guo}}, \bibinfo {author} {\bibfnamefont
  {Z.}~\bibnamefont {Xu}}, \bibinfo {author} {\bibfnamefont {J.}~\bibnamefont
  {Wang}}, \bibinfo {author} {\bibfnamefont {X.~H.}\ \bibnamefont {Chen}},\
  and\ \bibinfo {author} {\bibfnamefont {Y.}~\bibnamefont {Zhang}},\ }\bibfield
   {title} {\bibinfo {title} {Quantum anomalous {{Hall}} effect in intrinsic
  magnetic topological insulator {{MnBi2Te4}}},\ }\href
  {https://doi.org/10.1126/science.aax8156} {\bibfield  {journal} {\bibinfo
  {journal} {Science}\ }\textbf {\bibinfo {volume} {367}},\ \bibinfo {pages}
  {895} (\bibinfo {year} {2020})}\BibitemShut {NoStop}%
\bibitem [{\citenamefont {Liu}\ \emph {et~al.}(2020{\natexlab{a}})\citenamefont
  {Liu}, \citenamefont {Wang}, \citenamefont {Li}, \citenamefont {Wu},
  \citenamefont {Li}, \citenamefont {Li}, \citenamefont {He}, \citenamefont
  {Xu}, \citenamefont {Zhang},\ and\ \citenamefont {Wang}}]{liu2020robust}%
  \BibitemOpen
  \bibfield  {author} {\bibinfo {author} {\bibfnamefont {C.}~\bibnamefont
  {Liu}}, \bibinfo {author} {\bibfnamefont {Y.}~\bibnamefont {Wang}}, \bibinfo
  {author} {\bibfnamefont {H.}~\bibnamefont {Li}}, \bibinfo {author}
  {\bibfnamefont {Y.}~\bibnamefont {Wu}}, \bibinfo {author} {\bibfnamefont
  {Y.}~\bibnamefont {Li}}, \bibinfo {author} {\bibfnamefont {J.}~\bibnamefont
  {Li}}, \bibinfo {author} {\bibfnamefont {K.}~\bibnamefont {He}}, \bibinfo
  {author} {\bibfnamefont {Y.}~\bibnamefont {Xu}}, \bibinfo {author}
  {\bibfnamefont {J.}~\bibnamefont {Zhang}},\ and\ \bibinfo {author}
  {\bibfnamefont {Y.}~\bibnamefont {Wang}},\ }\bibfield  {title} {\bibinfo
  {title} {Robust axion insulator and {{Chern}} insulator phases in a
  two-dimensional antiferromagnetic topological insulator},\ }\href
  {https://doi.org/10.1038/s41563-019-0573-3} {\bibfield  {journal} {\bibinfo
  {journal} {Nat. Mater.}\ }\textbf {\bibinfo {volume} {19}},\ \bibinfo {pages}
  {522} (\bibinfo {year} {2020}{\natexlab{a}})}\BibitemShut {NoStop}%
\bibitem [{\citenamefont {Sharpe}\ \emph {et~al.}(2019)\citenamefont {Sharpe},
  \citenamefont {Fox}, \citenamefont {Barnard}, \citenamefont {Finney},
  \citenamefont {Watanabe}, \citenamefont {Taniguchi}, \citenamefont
  {Kastner},\ and\ \citenamefont {{Goldhaber-Gordon}}}]{sharpe2019emergent}%
  \BibitemOpen
  \bibfield  {author} {\bibinfo {author} {\bibfnamefont {A.~L.}\ \bibnamefont
  {Sharpe}}, \bibinfo {author} {\bibfnamefont {E.~J.}\ \bibnamefont {Fox}},
  \bibinfo {author} {\bibfnamefont {A.~W.}\ \bibnamefont {Barnard}}, \bibinfo
  {author} {\bibfnamefont {J.}~\bibnamefont {Finney}}, \bibinfo {author}
  {\bibfnamefont {K.}~\bibnamefont {Watanabe}}, \bibinfo {author}
  {\bibfnamefont {T.}~\bibnamefont {Taniguchi}}, \bibinfo {author}
  {\bibfnamefont {M.~A.}\ \bibnamefont {Kastner}},\ and\ \bibinfo {author}
  {\bibfnamefont {D.}~\bibnamefont {{Goldhaber-Gordon}}},\ }\bibfield  {title}
  {\bibinfo {title} {Emergent ferromagnetism near three-quarters filling in
  twisted bilayer graphene},\ }\href {https://doi.org/10.1126/science.aaw3780}
  {\bibfield  {journal} {\bibinfo  {journal} {Science}\ }\textbf {\bibinfo
  {volume} {365}},\ \bibinfo {pages} {605} (\bibinfo {year}
  {2019})}\BibitemShut {NoStop}%
\bibitem [{\citenamefont {Serlin}\ \emph {et~al.}(2020)\citenamefont {Serlin},
  \citenamefont {Tschirhart}, \citenamefont {Polshyn}, \citenamefont {Zhang},
  \citenamefont {Zhu}, \citenamefont {Watanabe}, \citenamefont {Taniguchi},
  \citenamefont {Balents},\ and\ \citenamefont {Young}}]{serlin2020intrinsic}%
  \BibitemOpen
  \bibfield  {author} {\bibinfo {author} {\bibfnamefont {M.}~\bibnamefont
  {Serlin}}, \bibinfo {author} {\bibfnamefont {C.~L.}\ \bibnamefont
  {Tschirhart}}, \bibinfo {author} {\bibfnamefont {H.}~\bibnamefont {Polshyn}},
  \bibinfo {author} {\bibfnamefont {Y.}~\bibnamefont {Zhang}}, \bibinfo
  {author} {\bibfnamefont {J.}~\bibnamefont {Zhu}}, \bibinfo {author}
  {\bibfnamefont {K.}~\bibnamefont {Watanabe}}, \bibinfo {author}
  {\bibfnamefont {T.}~\bibnamefont {Taniguchi}}, \bibinfo {author}
  {\bibfnamefont {L.}~\bibnamefont {Balents}},\ and\ \bibinfo {author}
  {\bibfnamefont {A.~F.}\ \bibnamefont {Young}},\ }\bibfield  {title} {\bibinfo
  {title} {Intrinsic quantized anomalous {{Hall}} effect in a moir\'e
  heterostructure},\ }\href {https://doi.org/10.1126/science.aay5533}
  {\bibfield  {journal} {\bibinfo  {journal} {Science}\ }\textbf {\bibinfo
  {volume} {367}},\ \bibinfo {pages} {900} (\bibinfo {year}
  {2020})}\BibitemShut {NoStop}%
\bibitem [{\citenamefont {Chen}\ \emph
  {et~al.}(2020{\natexlab{a}})\citenamefont {Chen}, \citenamefont {Sharpe},
  \citenamefont {Fox}, \citenamefont {Zhang}, \citenamefont {Wang},
  \citenamefont {Jiang}, \citenamefont {Lyu}, \citenamefont {Li}, \citenamefont
  {Watanabe}, \citenamefont {Taniguchi}, \citenamefont {Shi}, \citenamefont
  {Senthil}, \citenamefont {{Goldhaber-Gordon}}, \citenamefont {Zhang},\ and\
  \citenamefont {Wang}}]{chen2020tunable}%
  \BibitemOpen
  \bibfield  {author} {\bibinfo {author} {\bibfnamefont {G.}~\bibnamefont
  {Chen}}, \bibinfo {author} {\bibfnamefont {A.~L.}\ \bibnamefont {Sharpe}},
  \bibinfo {author} {\bibfnamefont {E.~J.}\ \bibnamefont {Fox}}, \bibinfo
  {author} {\bibfnamefont {Y.-H.}\ \bibnamefont {Zhang}}, \bibinfo {author}
  {\bibfnamefont {S.}~\bibnamefont {Wang}}, \bibinfo {author} {\bibfnamefont
  {L.}~\bibnamefont {Jiang}}, \bibinfo {author} {\bibfnamefont
  {B.}~\bibnamefont {Lyu}}, \bibinfo {author} {\bibfnamefont {H.}~\bibnamefont
  {Li}}, \bibinfo {author} {\bibfnamefont {K.}~\bibnamefont {Watanabe}},
  \bibinfo {author} {\bibfnamefont {T.}~\bibnamefont {Taniguchi}}, \bibinfo
  {author} {\bibfnamefont {Z.}~\bibnamefont {Shi}}, \bibinfo {author}
  {\bibfnamefont {T.}~\bibnamefont {Senthil}}, \bibinfo {author} {\bibfnamefont
  {D.}~\bibnamefont {{Goldhaber-Gordon}}}, \bibinfo {author} {\bibfnamefont
  {Y.}~\bibnamefont {Zhang}},\ and\ \bibinfo {author} {\bibfnamefont
  {F.}~\bibnamefont {Wang}},\ }\bibfield  {title} {\bibinfo {title} {Tunable
  correlated {{Chern}} insulator and ferromagnetism in a moir\'e
  superlattice},\ }\href {https://doi.org/10.1038/s41586-020-2049-7} {\bibfield
   {journal} {\bibinfo  {journal} {Nature}\ }\textbf {\bibinfo {volume}
  {579}},\ \bibinfo {pages} {56} (\bibinfo {year}
  {2020}{\natexlab{a}})}\BibitemShut {NoStop}%
\bibitem [{\citenamefont {Polshyn}\ \emph {et~al.}(2020)\citenamefont
  {Polshyn}, \citenamefont {Zhu}, \citenamefont {Kumar}, \citenamefont {Zhang},
  \citenamefont {Yang}, \citenamefont {Tschirhart}, \citenamefont {Serlin},
  \citenamefont {Watanabe}, \citenamefont {Taniguchi}, \citenamefont
  {MacDonald},\ and\ \citenamefont {Young}}]{polshyn2020electrical}%
  \BibitemOpen
  \bibfield  {author} {\bibinfo {author} {\bibfnamefont {H.}~\bibnamefont
  {Polshyn}}, \bibinfo {author} {\bibfnamefont {J.}~\bibnamefont {Zhu}},
  \bibinfo {author} {\bibfnamefont {M.~A.}\ \bibnamefont {Kumar}}, \bibinfo
  {author} {\bibfnamefont {Y.}~\bibnamefont {Zhang}}, \bibinfo {author}
  {\bibfnamefont {F.}~\bibnamefont {Yang}}, \bibinfo {author} {\bibfnamefont
  {C.~L.}\ \bibnamefont {Tschirhart}}, \bibinfo {author} {\bibfnamefont
  {M.}~\bibnamefont {Serlin}}, \bibinfo {author} {\bibfnamefont
  {K.}~\bibnamefont {Watanabe}}, \bibinfo {author} {\bibfnamefont
  {T.}~\bibnamefont {Taniguchi}}, \bibinfo {author} {\bibfnamefont {A.~H.}\
  \bibnamefont {MacDonald}},\ and\ \bibinfo {author} {\bibfnamefont {A.~F.}\
  \bibnamefont {Young}},\ }\bibfield  {title} {\bibinfo {title} {Electrical
  switching of magnetic order in an orbital {{Chern}} insulator},\ }\href
  {https://doi.org/10.1038/s41586-020-2963-8} {\bibfield  {journal} {\bibinfo
  {journal} {Nature}\ }\textbf {\bibinfo {volume} {588}},\ \bibinfo {pages}
  {66} (\bibinfo {year} {2020})}\BibitemShut {NoStop}%
\bibitem [{\citenamefont {Polshyn}\ \emph {et~al.}(2022)\citenamefont
  {Polshyn}, \citenamefont {Zhang}, \citenamefont {Kumar}, \citenamefont
  {Soejima}, \citenamefont {Ledwith}, \citenamefont {Watanabe}, \citenamefont
  {Taniguchi}, \citenamefont {Vishwanath}, \citenamefont {Zaletel},\ and\
  \citenamefont {Young}}]{polshyn2022topological}%
  \BibitemOpen
  \bibfield  {author} {\bibinfo {author} {\bibfnamefont {H.}~\bibnamefont
  {Polshyn}}, \bibinfo {author} {\bibfnamefont {Y.}~\bibnamefont {Zhang}},
  \bibinfo {author} {\bibfnamefont {M.~A.}\ \bibnamefont {Kumar}}, \bibinfo
  {author} {\bibfnamefont {T.}~\bibnamefont {Soejima}}, \bibinfo {author}
  {\bibfnamefont {P.}~\bibnamefont {Ledwith}}, \bibinfo {author} {\bibfnamefont
  {K.}~\bibnamefont {Watanabe}}, \bibinfo {author} {\bibfnamefont
  {T.}~\bibnamefont {Taniguchi}}, \bibinfo {author} {\bibfnamefont
  {A.}~\bibnamefont {Vishwanath}}, \bibinfo {author} {\bibfnamefont {M.~P.}\
  \bibnamefont {Zaletel}},\ and\ \bibinfo {author} {\bibfnamefont {A.~F.}\
  \bibnamefont {Young}},\ }\bibfield  {title} {\bibinfo {title} {Topological
  charge density waves at half-integer filling of a moir\'e superlattice},\
  }\href {https://doi.org/10.1038/s41567-021-01418-6} {\bibfield  {journal}
  {\bibinfo  {journal} {Nat. Phys.}\ }\textbf {\bibinfo {volume} {18}},\
  \bibinfo {pages} {42} (\bibinfo {year} {2022})}\BibitemShut {NoStop}%
\bibitem [{\citenamefont {K{\"o}nig}\ \emph {et~al.}(2007)\citenamefont
  {K{\"o}nig}, \citenamefont {Wiedmann}, \citenamefont {Br{\"u}ne},
  \citenamefont {Roth}, \citenamefont {Buhmann}, \citenamefont {Molenkamp},
  \citenamefont {Qi},\ and\ \citenamefont {Zhang}}]{konig2007quantum}%
  \BibitemOpen
  \bibfield  {author} {\bibinfo {author} {\bibfnamefont {M.}~\bibnamefont
  {K{\"o}nig}}, \bibinfo {author} {\bibfnamefont {S.}~\bibnamefont {Wiedmann}},
  \bibinfo {author} {\bibfnamefont {C.}~\bibnamefont {Br{\"u}ne}}, \bibinfo
  {author} {\bibfnamefont {A.}~\bibnamefont {Roth}}, \bibinfo {author}
  {\bibfnamefont {H.}~\bibnamefont {Buhmann}}, \bibinfo {author} {\bibfnamefont
  {L.~W.}\ \bibnamefont {Molenkamp}}, \bibinfo {author} {\bibfnamefont {X.-L.}\
  \bibnamefont {Qi}},\ and\ \bibinfo {author} {\bibfnamefont {S.-C.}\
  \bibnamefont {Zhang}},\ }\bibfield  {title} {\bibinfo {title} {Quantum {{Spin
  Hall Insulator State}} in {{HgTe Quantum Wells}}},\ }\href
  {https://doi.org/10.1126/science.1148047} {\bibfield  {journal} {\bibinfo
  {journal} {Science}\ }\textbf {\bibinfo {volume} {318}},\ \bibinfo {pages}
  {766} (\bibinfo {year} {2007})}\BibitemShut {NoStop}%
\bibitem [{\citenamefont {Knez}\ \emph {et~al.}(2010)\citenamefont {Knez},
  \citenamefont {Du},\ and\ \citenamefont {Sullivan}}]{knez2010finite}%
  \BibitemOpen
  \bibfield  {author} {\bibinfo {author} {\bibfnamefont {I.}~\bibnamefont
  {Knez}}, \bibinfo {author} {\bibfnamefont {R.~R.}\ \bibnamefont {Du}},\ and\
  \bibinfo {author} {\bibfnamefont {G.}~\bibnamefont {Sullivan}},\ }\bibfield
  {title} {\bibinfo {title} {Finite conductivity in mesoscopic {{Hall}} bars of
  inverted {{InAs}}/{{GaSb}} quantum wells},\ }\href
  {https://doi.org/10.1103/PhysRevB.81.201301} {\bibfield  {journal} {\bibinfo
  {journal} {Phys. Rev. B}\ }\textbf {\bibinfo {volume} {81}},\ \bibinfo
  {pages} {201301} (\bibinfo {year} {2010})}\BibitemShut {NoStop}%
\bibitem [{\citenamefont {Knez}\ \emph {et~al.}(2011)\citenamefont {Knez},
  \citenamefont {Du},\ and\ \citenamefont {Sullivan}}]{knez2011evidence}%
  \BibitemOpen
  \bibfield  {author} {\bibinfo {author} {\bibfnamefont {I.}~\bibnamefont
  {Knez}}, \bibinfo {author} {\bibfnamefont {R.-R.}\ \bibnamefont {Du}},\ and\
  \bibinfo {author} {\bibfnamefont {G.}~\bibnamefont {Sullivan}},\ }\bibfield
  {title} {\bibinfo {title} {Evidence for {{Helical Edge Modes}} in
  {{Inverted}} {$\mathrm{InAs}/\mathrm{GaSb}$} {{Quantum Wells}}},\ }\href
  {https://doi.org/10.1103/PhysRevLett.107.136603} {\bibfield  {journal}
  {\bibinfo  {journal} {Phys. Rev. Lett.}\ }\textbf {\bibinfo {volume} {107}},\
  \bibinfo {pages} {136603} (\bibinfo {year} {2011})}\BibitemShut {NoStop}%
\bibitem [{\citenamefont {Fei}\ \emph {et~al.}(2017)\citenamefont {Fei},
  \citenamefont {Palomaki}, \citenamefont {Wu}, \citenamefont {Zhao},
  \citenamefont {Cai}, \citenamefont {Sun}, \citenamefont {Nguyen},
  \citenamefont {Finney}, \citenamefont {Xu},\ and\ \citenamefont
  {Cobden}}]{fei2017edge}%
  \BibitemOpen
  \bibfield  {author} {\bibinfo {author} {\bibfnamefont {Z.}~\bibnamefont
  {Fei}}, \bibinfo {author} {\bibfnamefont {T.}~\bibnamefont {Palomaki}},
  \bibinfo {author} {\bibfnamefont {S.}~\bibnamefont {Wu}}, \bibinfo {author}
  {\bibfnamefont {W.}~\bibnamefont {Zhao}}, \bibinfo {author} {\bibfnamefont
  {X.}~\bibnamefont {Cai}}, \bibinfo {author} {\bibfnamefont {B.}~\bibnamefont
  {Sun}}, \bibinfo {author} {\bibfnamefont {P.}~\bibnamefont {Nguyen}},
  \bibinfo {author} {\bibfnamefont {J.}~\bibnamefont {Finney}}, \bibinfo
  {author} {\bibfnamefont {X.}~\bibnamefont {Xu}},\ and\ \bibinfo {author}
  {\bibfnamefont {D.~H.}\ \bibnamefont {Cobden}},\ }\bibfield  {title}
  {\bibinfo {title} {Edge conduction in monolayer {{WTe2}}},\ }\href
  {https://doi.org/10.1038/nphys4091} {\bibfield  {journal} {\bibinfo
  {journal} {Nature Phys}\ }\textbf {\bibinfo {volume} {13}},\ \bibinfo {pages}
  {677} (\bibinfo {year} {2017})}\BibitemShut {NoStop}%
\bibitem [{\citenamefont {Wu}\ \emph {et~al.}(2018{\natexlab{a}})\citenamefont
  {Wu}, \citenamefont {Fatemi}, \citenamefont {Gibson}, \citenamefont
  {Watanabe}, \citenamefont {Taniguchi}, \citenamefont {Cava},\ and\
  \citenamefont {{Jarillo-Herrero}}}]{wu2018observation}%
  \BibitemOpen
  \bibfield  {author} {\bibinfo {author} {\bibfnamefont {S.}~\bibnamefont
  {Wu}}, \bibinfo {author} {\bibfnamefont {V.}~\bibnamefont {Fatemi}}, \bibinfo
  {author} {\bibfnamefont {Q.~D.}\ \bibnamefont {Gibson}}, \bibinfo {author}
  {\bibfnamefont {K.}~\bibnamefont {Watanabe}}, \bibinfo {author}
  {\bibfnamefont {T.}~\bibnamefont {Taniguchi}}, \bibinfo {author}
  {\bibfnamefont {R.~J.}\ \bibnamefont {Cava}},\ and\ \bibinfo {author}
  {\bibfnamefont {P.}~\bibnamefont {{Jarillo-Herrero}}},\ }\bibfield  {title}
  {\bibinfo {title} {Observation of the quantum spin {{Hall}} effect up to 100
  kelvin in a monolayer crystal},\ }\href
  {https://doi.org/10.1126/science.aan6003} {\bibfield  {journal} {\bibinfo
  {journal} {Science}\ }\textbf {\bibinfo {volume} {359}},\ \bibinfo {pages}
  {76} (\bibinfo {year} {2018}{\natexlab{a}})}\BibitemShut {NoStop}%
\bibitem [{\citenamefont {Cao}\ \emph {et~al.}(2018{\natexlab{a}})\citenamefont
  {Cao}, \citenamefont {Fatemi}, \citenamefont {Demir}, \citenamefont {Fang},
  \citenamefont {Tomarken}, \citenamefont {Luo}, \citenamefont
  {{Sanchez-Yamagishi}}, \citenamefont {Watanabe}, \citenamefont {Taniguchi},
  \citenamefont {Kaxiras}, \citenamefont {Ashoori},\ and\ \citenamefont
  {{Jarillo-Herrero}}}]{cao2018correlated}%
  \BibitemOpen
  \bibfield  {author} {\bibinfo {author} {\bibfnamefont {Y.}~\bibnamefont
  {Cao}}, \bibinfo {author} {\bibfnamefont {V.}~\bibnamefont {Fatemi}},
  \bibinfo {author} {\bibfnamefont {A.}~\bibnamefont {Demir}}, \bibinfo
  {author} {\bibfnamefont {S.}~\bibnamefont {Fang}}, \bibinfo {author}
  {\bibfnamefont {S.~L.}\ \bibnamefont {Tomarken}}, \bibinfo {author}
  {\bibfnamefont {J.~Y.}\ \bibnamefont {Luo}}, \bibinfo {author} {\bibfnamefont
  {J.~D.}\ \bibnamefont {{Sanchez-Yamagishi}}}, \bibinfo {author}
  {\bibfnamefont {K.}~\bibnamefont {Watanabe}}, \bibinfo {author}
  {\bibfnamefont {T.}~\bibnamefont {Taniguchi}}, \bibinfo {author}
  {\bibfnamefont {E.}~\bibnamefont {Kaxiras}}, \bibinfo {author} {\bibfnamefont
  {R.~C.}\ \bibnamefont {Ashoori}},\ and\ \bibinfo {author} {\bibfnamefont
  {P.}~\bibnamefont {{Jarillo-Herrero}}},\ }\bibfield  {title} {\bibinfo
  {title} {Correlated insulator behaviour at half-filling in magic-angle
  graphene superlattices},\ }\href {https://doi.org/10.1038/nature26154}
  {\bibfield  {journal} {\bibinfo  {journal} {Nature}\ }\textbf {\bibinfo
  {volume} {556}},\ \bibinfo {pages} {80} (\bibinfo {year}
  {2018}{\natexlab{a}})}\BibitemShut {NoStop}%
\bibitem [{\citenamefont {Cao}\ \emph {et~al.}(2018{\natexlab{b}})\citenamefont
  {Cao}, \citenamefont {Fatemi}, \citenamefont {Fang}, \citenamefont
  {Watanabe}, \citenamefont {Taniguchi}, \citenamefont {Kaxiras},\ and\
  \citenamefont {{Jarillo-Herrero}}}]{cao2018unconventional}%
  \BibitemOpen
  \bibfield  {author} {\bibinfo {author} {\bibfnamefont {Y.}~\bibnamefont
  {Cao}}, \bibinfo {author} {\bibfnamefont {V.}~\bibnamefont {Fatemi}},
  \bibinfo {author} {\bibfnamefont {S.}~\bibnamefont {Fang}}, \bibinfo {author}
  {\bibfnamefont {K.}~\bibnamefont {Watanabe}}, \bibinfo {author}
  {\bibfnamefont {T.}~\bibnamefont {Taniguchi}}, \bibinfo {author}
  {\bibfnamefont {E.}~\bibnamefont {Kaxiras}},\ and\ \bibinfo {author}
  {\bibfnamefont {P.}~\bibnamefont {{Jarillo-Herrero}}},\ }\bibfield  {title}
  {\bibinfo {title} {Unconventional superconductivity in magic-angle graphene
  superlattices},\ }\href {https://doi.org/10.1038/nature26160} {\bibfield
  {journal} {\bibinfo  {journal} {Nature}\ }\textbf {\bibinfo {volume} {556}},\
  \bibinfo {pages} {43} (\bibinfo {year} {2018}{\natexlab{b}})}\BibitemShut
  {NoStop}%
\bibitem [{\citenamefont {Yankowitz}\ \emph {et~al.}(2019)\citenamefont
  {Yankowitz}, \citenamefont {Chen}, \citenamefont {Polshyn}, \citenamefont
  {Zhang}, \citenamefont {Watanabe}, \citenamefont {Taniguchi}, \citenamefont
  {Graf}, \citenamefont {Young},\ and\ \citenamefont
  {Dean}}]{yankowitz2019tuning}%
  \BibitemOpen
  \bibfield  {author} {\bibinfo {author} {\bibfnamefont {M.}~\bibnamefont
  {Yankowitz}}, \bibinfo {author} {\bibfnamefont {S.}~\bibnamefont {Chen}},
  \bibinfo {author} {\bibfnamefont {H.}~\bibnamefont {Polshyn}}, \bibinfo
  {author} {\bibfnamefont {Y.}~\bibnamefont {Zhang}}, \bibinfo {author}
  {\bibfnamefont {K.}~\bibnamefont {Watanabe}}, \bibinfo {author}
  {\bibfnamefont {T.}~\bibnamefont {Taniguchi}}, \bibinfo {author}
  {\bibfnamefont {D.}~\bibnamefont {Graf}}, \bibinfo {author} {\bibfnamefont
  {A.~F.}\ \bibnamefont {Young}},\ and\ \bibinfo {author} {\bibfnamefont
  {C.~R.}\ \bibnamefont {Dean}},\ }\bibfield  {title} {\bibinfo {title} {Tuning
  superconductivity in twisted bilayer graphene},\ }\href
  {https://doi.org/10.1126/science.aav1910} {\bibfield  {journal} {\bibinfo
  {journal} {Science}\ }\textbf {\bibinfo {volume} {363}},\ \bibinfo {pages}
  {1059} (\bibinfo {year} {2019})}\BibitemShut {NoStop}%
\bibitem [{\citenamefont {Chen}\ \emph
  {et~al.}(2019{\natexlab{a}})\citenamefont {Chen}, \citenamefont {Jiang},
  \citenamefont {Wu}, \citenamefont {Lyu}, \citenamefont {Li}, \citenamefont
  {Chittari}, \citenamefont {Watanabe}, \citenamefont {Taniguchi},
  \citenamefont {Shi}, \citenamefont {Jung}, \citenamefont {Zhang},\ and\
  \citenamefont {Wang}}]{chen2019evidence}%
  \BibitemOpen
  \bibfield  {author} {\bibinfo {author} {\bibfnamefont {G.}~\bibnamefont
  {Chen}}, \bibinfo {author} {\bibfnamefont {L.}~\bibnamefont {Jiang}},
  \bibinfo {author} {\bibfnamefont {S.}~\bibnamefont {Wu}}, \bibinfo {author}
  {\bibfnamefont {B.}~\bibnamefont {Lyu}}, \bibinfo {author} {\bibfnamefont
  {H.}~\bibnamefont {Li}}, \bibinfo {author} {\bibfnamefont {B.~L.}\
  \bibnamefont {Chittari}}, \bibinfo {author} {\bibfnamefont {K.}~\bibnamefont
  {Watanabe}}, \bibinfo {author} {\bibfnamefont {T.}~\bibnamefont {Taniguchi}},
  \bibinfo {author} {\bibfnamefont {Z.}~\bibnamefont {Shi}}, \bibinfo {author}
  {\bibfnamefont {J.}~\bibnamefont {Jung}}, \bibinfo {author} {\bibfnamefont
  {Y.}~\bibnamefont {Zhang}},\ and\ \bibinfo {author} {\bibfnamefont
  {F.}~\bibnamefont {Wang}},\ }\bibfield  {title} {\bibinfo {title} {Evidence
  of {{Gate-Tunable Mott Insulator}} in {{Trilayer Graphene-Boron Nitride
  Moir\'e Superlattice}}},\ }\href {https://doi.org/10.1038/s41567-018-0387-2}
  {\bibfield  {journal} {\bibinfo  {journal} {Nat. Phys.}\ }\textbf {\bibinfo
  {volume} {15}},\ \bibinfo {pages} {237} (\bibinfo {year}
  {2019}{\natexlab{a}})}\BibitemShut {NoStop}%
\bibitem [{\citenamefont {Cao}\ \emph {et~al.}(2020)\citenamefont {Cao},
  \citenamefont {{Rodan-Legrain}}, \citenamefont {{Rubies-Bigorda}},
  \citenamefont {Park}, \citenamefont {Watanabe}, \citenamefont {Taniguchi},\
  and\ \citenamefont {{Jarillo-Herrero}}}]{cao2020tunable}%
  \BibitemOpen
  \bibfield  {author} {\bibinfo {author} {\bibfnamefont {Y.}~\bibnamefont
  {Cao}}, \bibinfo {author} {\bibfnamefont {D.}~\bibnamefont
  {{Rodan-Legrain}}}, \bibinfo {author} {\bibfnamefont {O.}~\bibnamefont
  {{Rubies-Bigorda}}}, \bibinfo {author} {\bibfnamefont {J.~M.}\ \bibnamefont
  {Park}}, \bibinfo {author} {\bibfnamefont {K.}~\bibnamefont {Watanabe}},
  \bibinfo {author} {\bibfnamefont {T.}~\bibnamefont {Taniguchi}},\ and\
  \bibinfo {author} {\bibfnamefont {P.}~\bibnamefont {{Jarillo-Herrero}}},\
  }\bibfield  {title} {\bibinfo {title} {Tunable correlated states and
  spin-polarized phases in twisted bilayer\textendash bilayer graphene},\
  }\href {https://doi.org/10.1038/s41586-020-2260-6} {\bibfield  {journal}
  {\bibinfo  {journal} {Nature}\ }\textbf {\bibinfo {volume} {583}},\ \bibinfo
  {pages} {215} (\bibinfo {year} {2020})}\BibitemShut {NoStop}%
\bibitem [{\citenamefont {Chen}\ \emph
  {et~al.}(2020{\natexlab{b}})\citenamefont {Chen}, \citenamefont {He},
  \citenamefont {Zhang}, \citenamefont {Hsieh}, \citenamefont {Fei},
  \citenamefont {Watanabe}, \citenamefont {Taniguchi}, \citenamefont {Cobden},
  \citenamefont {Xu}, \citenamefont {Dean},\ and\ \citenamefont
  {Yankowitz}}]{chen2020electrically}%
  \BibitemOpen
  \bibfield  {author} {\bibinfo {author} {\bibfnamefont {S.}~\bibnamefont
  {Chen}}, \bibinfo {author} {\bibfnamefont {M.}~\bibnamefont {He}}, \bibinfo
  {author} {\bibfnamefont {Y.-H.}\ \bibnamefont {Zhang}}, \bibinfo {author}
  {\bibfnamefont {V.}~\bibnamefont {Hsieh}}, \bibinfo {author} {\bibfnamefont
  {Z.}~\bibnamefont {Fei}}, \bibinfo {author} {\bibfnamefont {K.}~\bibnamefont
  {Watanabe}}, \bibinfo {author} {\bibfnamefont {T.}~\bibnamefont {Taniguchi}},
  \bibinfo {author} {\bibfnamefont {D.~H.}\ \bibnamefont {Cobden}}, \bibinfo
  {author} {\bibfnamefont {X.}~\bibnamefont {Xu}}, \bibinfo {author}
  {\bibfnamefont {C.~R.}\ \bibnamefont {Dean}},\ and\ \bibinfo {author}
  {\bibfnamefont {M.}~\bibnamefont {Yankowitz}},\ }\bibfield  {title} {\bibinfo
  {title} {Electrically tunable correlated and topological states in twisted
  monolayer\textendash bilayer graphene},\ }\href
  {https://doi.org/10.1038/s41567-020-01062-6} {\bibfield  {journal} {\bibinfo
  {journal} {Nature Physics}\ }\textbf {\bibinfo {volume} {17}},\ \bibinfo
  {pages} {374} (\bibinfo {year} {2020}{\natexlab{b}})}\BibitemShut {NoStop}%
\bibitem [{\citenamefont {Wong}\ \emph {et~al.}(2020)\citenamefont {Wong},
  \citenamefont {Nuckolls}, \citenamefont {Oh}, \citenamefont {Lian},
  \citenamefont {Xie}, \citenamefont {Jeon}, \citenamefont {Watanabe},
  \citenamefont {Taniguchi}, \citenamefont {Bernevig},\ and\ \citenamefont
  {Yazdani}}]{wong2020cascade}%
  \BibitemOpen
  \bibfield  {author} {\bibinfo {author} {\bibfnamefont {D.}~\bibnamefont
  {Wong}}, \bibinfo {author} {\bibfnamefont {K.~P.}\ \bibnamefont {Nuckolls}},
  \bibinfo {author} {\bibfnamefont {M.}~\bibnamefont {Oh}}, \bibinfo {author}
  {\bibfnamefont {B.}~\bibnamefont {Lian}}, \bibinfo {author} {\bibfnamefont
  {Y.}~\bibnamefont {Xie}}, \bibinfo {author} {\bibfnamefont {S.}~\bibnamefont
  {Jeon}}, \bibinfo {author} {\bibfnamefont {K.}~\bibnamefont {Watanabe}},
  \bibinfo {author} {\bibfnamefont {T.}~\bibnamefont {Taniguchi}}, \bibinfo
  {author} {\bibfnamefont {B.~A.}\ \bibnamefont {Bernevig}},\ and\ \bibinfo
  {author} {\bibfnamefont {A.}~\bibnamefont {Yazdani}},\ }\bibfield  {title}
  {\bibinfo {title} {Cascade of electronic transitions in magic-angle twisted
  bilayer graphene},\ }\href {https://doi.org/10.1038/s41586-020-2339-0}
  {\bibfield  {journal} {\bibinfo  {journal} {Nature}\ }\textbf {\bibinfo
  {volume} {582}},\ \bibinfo {pages} {198} (\bibinfo {year}
  {2020})}\BibitemShut {NoStop}%
\bibitem [{\citenamefont {Lu}\ \emph {et~al.}(2019)\citenamefont {Lu},
  \citenamefont {Stepanov}, \citenamefont {Yang}, \citenamefont {Xie},
  \citenamefont {Aamir}, \citenamefont {Das}, \citenamefont {Urgell},
  \citenamefont {Watanabe}, \citenamefont {Taniguchi}, \citenamefont {Zhang},
  \citenamefont {Bachtold}, \citenamefont {MacDonald},\ and\ \citenamefont
  {Efetov}}]{lu2019superconductors}%
  \BibitemOpen
  \bibfield  {author} {\bibinfo {author} {\bibfnamefont {X.}~\bibnamefont
  {Lu}}, \bibinfo {author} {\bibfnamefont {P.}~\bibnamefont {Stepanov}},
  \bibinfo {author} {\bibfnamefont {W.}~\bibnamefont {Yang}}, \bibinfo {author}
  {\bibfnamefont {M.}~\bibnamefont {Xie}}, \bibinfo {author} {\bibfnamefont
  {M.~A.}\ \bibnamefont {Aamir}}, \bibinfo {author} {\bibfnamefont
  {I.}~\bibnamefont {Das}}, \bibinfo {author} {\bibfnamefont {C.}~\bibnamefont
  {Urgell}}, \bibinfo {author} {\bibfnamefont {K.}~\bibnamefont {Watanabe}},
  \bibinfo {author} {\bibfnamefont {T.}~\bibnamefont {Taniguchi}}, \bibinfo
  {author} {\bibfnamefont {G.}~\bibnamefont {Zhang}}, \bibinfo {author}
  {\bibfnamefont {A.}~\bibnamefont {Bachtold}}, \bibinfo {author}
  {\bibfnamefont {A.~H.}\ \bibnamefont {MacDonald}},\ and\ \bibinfo {author}
  {\bibfnamefont {D.~K.}\ \bibnamefont {Efetov}},\ }\bibfield  {title}
  {\bibinfo {title} {Superconductors, orbital magnets and correlated states in
  magic-angle bilayer graphene},\ }\href
  {https://doi.org/10.1038/s41586-019-1695-0} {\bibfield  {journal} {\bibinfo
  {journal} {Nature}\ }\textbf {\bibinfo {volume} {574}},\ \bibinfo {pages}
  {653} (\bibinfo {year} {2019})}\BibitemShut {NoStop}%
\bibitem [{\citenamefont {Shen}\ \emph {et~al.}(2020)\citenamefont {Shen},
  \citenamefont {Chu}, \citenamefont {Wu}, \citenamefont {Li}, \citenamefont
  {Wang}, \citenamefont {Zhao}, \citenamefont {Tang}, \citenamefont {Liu},
  \citenamefont {Tian}, \citenamefont {Watanabe}, \citenamefont {Taniguchi},
  \citenamefont {Yang}, \citenamefont {Meng}, \citenamefont {Shi},
  \citenamefont {Yazyev},\ and\ \citenamefont {Zhang}}]{shen2020correlated}%
  \BibitemOpen
  \bibfield  {author} {\bibinfo {author} {\bibfnamefont {C.}~\bibnamefont
  {Shen}}, \bibinfo {author} {\bibfnamefont {Y.}~\bibnamefont {Chu}}, \bibinfo
  {author} {\bibfnamefont {Q.}~\bibnamefont {Wu}}, \bibinfo {author}
  {\bibfnamefont {N.}~\bibnamefont {Li}}, \bibinfo {author} {\bibfnamefont
  {S.}~\bibnamefont {Wang}}, \bibinfo {author} {\bibfnamefont {Y.}~\bibnamefont
  {Zhao}}, \bibinfo {author} {\bibfnamefont {J.}~\bibnamefont {Tang}}, \bibinfo
  {author} {\bibfnamefont {J.}~\bibnamefont {Liu}}, \bibinfo {author}
  {\bibfnamefont {J.}~\bibnamefont {Tian}}, \bibinfo {author} {\bibfnamefont
  {K.}~\bibnamefont {Watanabe}}, \bibinfo {author} {\bibfnamefont
  {T.}~\bibnamefont {Taniguchi}}, \bibinfo {author} {\bibfnamefont
  {R.}~\bibnamefont {Yang}}, \bibinfo {author} {\bibfnamefont {Z.~Y.}\
  \bibnamefont {Meng}}, \bibinfo {author} {\bibfnamefont {D.}~\bibnamefont
  {Shi}}, \bibinfo {author} {\bibfnamefont {O.~V.}\ \bibnamefont {Yazyev}},\
  and\ \bibinfo {author} {\bibfnamefont {G.}~\bibnamefont {Zhang}},\ }\bibfield
   {title} {\bibinfo {title} {Correlated states in twisted double bilayer
  graphene},\ }\href {https://doi.org/10.1038/s41567-020-0825-9} {\bibfield
  {journal} {\bibinfo  {journal} {Nat. Phys.}\ }\textbf {\bibinfo {volume}
  {16}},\ \bibinfo {pages} {520} (\bibinfo {year} {2020})}\BibitemShut
  {NoStop}%
\bibitem [{\citenamefont {Choi}\ \emph {et~al.}(2019)\citenamefont {Choi},
  \citenamefont {Kemmer}, \citenamefont {Peng}, \citenamefont {Thomson},
  \citenamefont {Arora}, \citenamefont {Polski}, \citenamefont {Zhang},
  \citenamefont {Ren}, \citenamefont {Alicea}, \citenamefont {Refael},
  \citenamefont {{von Oppen}}, \citenamefont {Watanabe}, \citenamefont
  {Taniguchi},\ and\ \citenamefont {{Nadj-Perge}}}]{choi2019electronic}%
  \BibitemOpen
  \bibfield  {author} {\bibinfo {author} {\bibfnamefont {Y.}~\bibnamefont
  {Choi}}, \bibinfo {author} {\bibfnamefont {J.}~\bibnamefont {Kemmer}},
  \bibinfo {author} {\bibfnamefont {Y.}~\bibnamefont {Peng}}, \bibinfo {author}
  {\bibfnamefont {A.}~\bibnamefont {Thomson}}, \bibinfo {author} {\bibfnamefont
  {H.}~\bibnamefont {Arora}}, \bibinfo {author} {\bibfnamefont
  {R.}~\bibnamefont {Polski}}, \bibinfo {author} {\bibfnamefont
  {Y.}~\bibnamefont {Zhang}}, \bibinfo {author} {\bibfnamefont
  {H.}~\bibnamefont {Ren}}, \bibinfo {author} {\bibfnamefont {J.}~\bibnamefont
  {Alicea}}, \bibinfo {author} {\bibfnamefont {G.}~\bibnamefont {Refael}},
  \bibinfo {author} {\bibfnamefont {F.}~\bibnamefont {{von Oppen}}}, \bibinfo
  {author} {\bibfnamefont {K.}~\bibnamefont {Watanabe}}, \bibinfo {author}
  {\bibfnamefont {T.}~\bibnamefont {Taniguchi}},\ and\ \bibinfo {author}
  {\bibfnamefont {S.}~\bibnamefont {{Nadj-Perge}}},\ }\bibfield  {title}
  {\bibinfo {title} {Electronic correlations in twisted bilayer graphene near
  the magic angle},\ }\href {https://doi.org/10.1038/s41567-019-0606-5}
  {\bibfield  {journal} {\bibinfo  {journal} {Nat. Phys.}\ }\textbf {\bibinfo
  {volume} {15}},\ \bibinfo {pages} {1174} (\bibinfo {year}
  {2019})}\BibitemShut {NoStop}%
\bibitem [{\citenamefont {Xie}\ \emph {et~al.}(2019)\citenamefont {Xie},
  \citenamefont {Lian}, \citenamefont {J{\"a}ck}, \citenamefont {Liu},
  \citenamefont {Chiu}, \citenamefont {Watanabe}, \citenamefont {Taniguchi},
  \citenamefont {Bernevig},\ and\ \citenamefont
  {Yazdani}}]{xie2019spectroscopic}%
  \BibitemOpen
  \bibfield  {author} {\bibinfo {author} {\bibfnamefont {Y.}~\bibnamefont
  {Xie}}, \bibinfo {author} {\bibfnamefont {B.}~\bibnamefont {Lian}}, \bibinfo
  {author} {\bibfnamefont {B.}~\bibnamefont {J{\"a}ck}}, \bibinfo {author}
  {\bibfnamefont {X.}~\bibnamefont {Liu}}, \bibinfo {author} {\bibfnamefont
  {C.-L.}\ \bibnamefont {Chiu}}, \bibinfo {author} {\bibfnamefont
  {K.}~\bibnamefont {Watanabe}}, \bibinfo {author} {\bibfnamefont
  {T.}~\bibnamefont {Taniguchi}}, \bibinfo {author} {\bibfnamefont {B.~A.}\
  \bibnamefont {Bernevig}},\ and\ \bibinfo {author} {\bibfnamefont
  {A.}~\bibnamefont {Yazdani}},\ }\bibfield  {title} {\bibinfo {title}
  {Spectroscopic signatures of many-body correlations in magic-angle twisted
  bilayer graphene},\ }\href {https://doi.org/10.1038/s41586-019-1422-x}
  {\bibfield  {journal} {\bibinfo  {journal} {Nature}\ }\textbf {\bibinfo
  {volume} {572}},\ \bibinfo {pages} {101} (\bibinfo {year}
  {2019})}\BibitemShut {NoStop}%
\bibitem [{\citenamefont {Chen}\ \emph
  {et~al.}(2019{\natexlab{b}})\citenamefont {Chen}, \citenamefont {Sharpe},
  \citenamefont {Gallagher}, \citenamefont {Rosen}, \citenamefont {Fox},
  \citenamefont {Jiang}, \citenamefont {Lyu}, \citenamefont {Li}, \citenamefont
  {Watanabe}, \citenamefont {Taniguchi}, \citenamefont {Jung}, \citenamefont
  {Shi}, \citenamefont {{Goldhaber-Gordon}}, \citenamefont {Zhang},\ and\
  \citenamefont {Wang}}]{chen2019signatures}%
  \BibitemOpen
  \bibfield  {author} {\bibinfo {author} {\bibfnamefont {G.}~\bibnamefont
  {Chen}}, \bibinfo {author} {\bibfnamefont {A.~L.}\ \bibnamefont {Sharpe}},
  \bibinfo {author} {\bibfnamefont {P.}~\bibnamefont {Gallagher}}, \bibinfo
  {author} {\bibfnamefont {I.~T.}\ \bibnamefont {Rosen}}, \bibinfo {author}
  {\bibfnamefont {E.~J.}\ \bibnamefont {Fox}}, \bibinfo {author} {\bibfnamefont
  {L.}~\bibnamefont {Jiang}}, \bibinfo {author} {\bibfnamefont
  {B.}~\bibnamefont {Lyu}}, \bibinfo {author} {\bibfnamefont {H.}~\bibnamefont
  {Li}}, \bibinfo {author} {\bibfnamefont {K.}~\bibnamefont {Watanabe}},
  \bibinfo {author} {\bibfnamefont {T.}~\bibnamefont {Taniguchi}}, \bibinfo
  {author} {\bibfnamefont {J.}~\bibnamefont {Jung}}, \bibinfo {author}
  {\bibfnamefont {Z.}~\bibnamefont {Shi}}, \bibinfo {author} {\bibfnamefont
  {D.}~\bibnamefont {{Goldhaber-Gordon}}}, \bibinfo {author} {\bibfnamefont
  {Y.}~\bibnamefont {Zhang}},\ and\ \bibinfo {author} {\bibfnamefont
  {F.}~\bibnamefont {Wang}},\ }\bibfield  {title} {\bibinfo {title} {Signatures
  of tunable superconductivity in a trilayer graphene moir\'e superlattice},\
  }\href {https://doi.org/10.1038/s41586-019-1393-y} {\bibfield  {journal}
  {\bibinfo  {journal} {Nature}\ }\textbf {\bibinfo {volume} {572}},\ \bibinfo
  {pages} {215} (\bibinfo {year} {2019}{\natexlab{b}})}\BibitemShut {NoStop}%
\bibitem [{\citenamefont {Liu}\ \emph {et~al.}(2020{\natexlab{b}})\citenamefont
  {Liu}, \citenamefont {Hao}, \citenamefont {Khalaf}, \citenamefont {Lee},
  \citenamefont {Ronen}, \citenamefont {Yoo}, \citenamefont {Haei~Najafabadi},
  \citenamefont {Watanabe}, \citenamefont {Taniguchi}, \citenamefont
  {Vishwanath},\ and\ \citenamefont {Kim}}]{liu2020tunable}%
  \BibitemOpen
  \bibfield  {author} {\bibinfo {author} {\bibfnamefont {X.}~\bibnamefont
  {Liu}}, \bibinfo {author} {\bibfnamefont {Z.}~\bibnamefont {Hao}}, \bibinfo
  {author} {\bibfnamefont {E.}~\bibnamefont {Khalaf}}, \bibinfo {author}
  {\bibfnamefont {J.~Y.}\ \bibnamefont {Lee}}, \bibinfo {author} {\bibfnamefont
  {Y.}~\bibnamefont {Ronen}}, \bibinfo {author} {\bibfnamefont
  {H.}~\bibnamefont {Yoo}}, \bibinfo {author} {\bibfnamefont {D.}~\bibnamefont
  {Haei~Najafabadi}}, \bibinfo {author} {\bibfnamefont {K.}~\bibnamefont
  {Watanabe}}, \bibinfo {author} {\bibfnamefont {T.}~\bibnamefont {Taniguchi}},
  \bibinfo {author} {\bibfnamefont {A.}~\bibnamefont {Vishwanath}},\ and\
  \bibinfo {author} {\bibfnamefont {P.}~\bibnamefont {Kim}},\ }\bibfield
  {title} {\bibinfo {title} {Tunable spin-polarized correlated states in
  twisted double bilayer graphene},\ }\href
  {https://doi.org/10.1038/s41586-020-2458-7} {\bibfield  {journal} {\bibinfo
  {journal} {Nature}\ }\textbf {\bibinfo {volume} {583}},\ \bibinfo {pages}
  {221} (\bibinfo {year} {2020}{\natexlab{b}})}\BibitemShut {NoStop}%
\bibitem [{\citenamefont {Wu}\ \emph {et~al.}(2019)\citenamefont {Wu},
  \citenamefont {Lovorn}, \citenamefont {Tutuc}, \citenamefont {Martin},\ and\
  \citenamefont {MacDonald}}]{wu2019topological}%
  \BibitemOpen
  \bibfield  {author} {\bibinfo {author} {\bibfnamefont {F.}~\bibnamefont
  {Wu}}, \bibinfo {author} {\bibfnamefont {T.}~\bibnamefont {Lovorn}}, \bibinfo
  {author} {\bibfnamefont {E.}~\bibnamefont {Tutuc}}, \bibinfo {author}
  {\bibfnamefont {I.}~\bibnamefont {Martin}},\ and\ \bibinfo {author}
  {\bibfnamefont {A.~H.}\ \bibnamefont {MacDonald}},\ }\bibfield  {title}
  {\bibinfo {title} {Topological insulators in twisted transition metal
  dichalcogenide homobilayers},\ }\href
  {https://doi.org/10.1103/PhysRevLett.122.086402} {\bibfield  {journal}
  {\bibinfo  {journal} {Phys. Rev. Lett.}\ }\textbf {\bibinfo {volume} {122}},\
  \bibinfo {pages} {086402} (\bibinfo {year} {2019})}\BibitemShut {NoStop}%
\bibitem [{\citenamefont {Pan}\ \emph {et~al.}(2020{\natexlab{a}})\citenamefont
  {Pan}, \citenamefont {Wu},\ and\ \citenamefont {Das~Sarma}}]{pan2020band}%
  \BibitemOpen
  \bibfield  {author} {\bibinfo {author} {\bibfnamefont {H.}~\bibnamefont
  {Pan}}, \bibinfo {author} {\bibfnamefont {F.}~\bibnamefont {Wu}},\ and\
  \bibinfo {author} {\bibfnamefont {S.}~\bibnamefont {Das~Sarma}},\ }\bibfield
  {title} {\bibinfo {title} {Band topology, {{Hubbard}} model, {{Heisenberg}}
  model, and {{Dzyaloshinskii-Moriya}} interaction in twisted bilayer
  {${\mathrm{WSe}}_{2}$}},\ }\href
  {https://doi.org/10.1103/PhysRevResearch.2.033087} {\bibfield  {journal}
  {\bibinfo  {journal} {Phys. Rev. Research}\ }\textbf {\bibinfo {volume}
  {2}},\ \bibinfo {pages} {033087} (\bibinfo {year}
  {2020}{\natexlab{a}})}\BibitemShut {NoStop}%
\bibitem [{\citenamefont {Devakul}\ \emph {et~al.}(2021)\citenamefont
  {Devakul}, \citenamefont {Cr{\'e}pel}, \citenamefont {Zhang},\ and\
  \citenamefont {Fu}}]{devakul2021magic}%
  \BibitemOpen
  \bibfield  {author} {\bibinfo {author} {\bibfnamefont {T.}~\bibnamefont
  {Devakul}}, \bibinfo {author} {\bibfnamefont {V.}~\bibnamefont {Cr{\'e}pel}},
  \bibinfo {author} {\bibfnamefont {Y.}~\bibnamefont {Zhang}},\ and\ \bibinfo
  {author} {\bibfnamefont {L.}~\bibnamefont {Fu}},\ }\bibfield  {title}
  {\bibinfo {title} {Magic in twisted transition metal dichalcogenide
  bilayers},\ }\href {https://doi.org/10.1038/s41467-021-27042-9} {\bibfield
  {journal} {\bibinfo  {journal} {Nat Commun}\ }\textbf {\bibinfo {volume}
  {12}},\ \bibinfo {pages} {6730} (\bibinfo {year} {2021})}\BibitemShut
  {NoStop}%
\bibitem [{\citenamefont {Li}\ \emph {et~al.}(2021{\natexlab{a}})\citenamefont
  {Li}, \citenamefont {Jiang}, \citenamefont {Shen}, \citenamefont {Zhang},
  \citenamefont {Li}, \citenamefont {Tao}, \citenamefont {Devakul},
  \citenamefont {Watanabe}, \citenamefont {Taniguchi}, \citenamefont {Fu},
  \citenamefont {Shan},\ and\ \citenamefont {Mak}}]{li2021quantum}%
  \BibitemOpen
  \bibfield  {author} {\bibinfo {author} {\bibfnamefont {T.}~\bibnamefont
  {Li}}, \bibinfo {author} {\bibfnamefont {S.}~\bibnamefont {Jiang}}, \bibinfo
  {author} {\bibfnamefont {B.}~\bibnamefont {Shen}}, \bibinfo {author}
  {\bibfnamefont {Y.}~\bibnamefont {Zhang}}, \bibinfo {author} {\bibfnamefont
  {L.}~\bibnamefont {Li}}, \bibinfo {author} {\bibfnamefont {Z.}~\bibnamefont
  {Tao}}, \bibinfo {author} {\bibfnamefont {T.}~\bibnamefont {Devakul}},
  \bibinfo {author} {\bibfnamefont {K.}~\bibnamefont {Watanabe}}, \bibinfo
  {author} {\bibfnamefont {T.}~\bibnamefont {Taniguchi}}, \bibinfo {author}
  {\bibfnamefont {L.}~\bibnamefont {Fu}}, \bibinfo {author} {\bibfnamefont
  {J.}~\bibnamefont {Shan}},\ and\ \bibinfo {author} {\bibfnamefont {K.~F.}\
  \bibnamefont {Mak}},\ }\bibfield  {title} {\bibinfo {title} {Quantum
  anomalous {{Hall}} effect from intertwined moir\'e bands},\ }\href
  {https://doi.org/10.1038/s41586-021-04171-1} {\bibfield  {journal} {\bibinfo
  {journal} {Nature}\ }\textbf {\bibinfo {volume} {600}},\ \bibinfo {pages}
  {641} (\bibinfo {year} {2021}{\natexlab{a}})}\BibitemShut {NoStop}%
\bibitem [{\citenamefont {Xie}\ \emph {et~al.}(2022)\citenamefont {Xie},
  \citenamefont {Zhang}, \citenamefont {Hu}, \citenamefont {Mak},\ and\
  \citenamefont {Law}}]{xie2022valleypolarized}%
  \BibitemOpen
  \bibfield  {author} {\bibinfo {author} {\bibfnamefont {Y.-M.}\ \bibnamefont
  {Xie}}, \bibinfo {author} {\bibfnamefont {C.-P.}\ \bibnamefont {Zhang}},
  \bibinfo {author} {\bibfnamefont {J.-X.}\ \bibnamefont {Hu}}, \bibinfo
  {author} {\bibfnamefont {K.~F.}\ \bibnamefont {Mak}},\ and\ \bibinfo {author}
  {\bibfnamefont {K.~T.}\ \bibnamefont {Law}},\ }\bibfield  {title} {\bibinfo
  {title} {Valley-{{Polarized Quantum Anomalous Hall State}} in {{Moir\'e}}
  {${\mathrm{MoTe}}_{2}/{\mathrm{WSe}}_{2}$} {{Heterobilayers}}},\ }\href
  {https://doi.org/10.1103/PhysRevLett.128.026402} {\bibfield  {journal}
  {\bibinfo  {journal} {Phys. Rev. Lett.}\ }\textbf {\bibinfo {volume} {128}},\
  \bibinfo {pages} {026402} (\bibinfo {year} {2022})}\BibitemShut {NoStop}%
\bibitem [{\citenamefont {Zhang}\ \emph
  {et~al.}(2021{\natexlab{a}})\citenamefont {Zhang}, \citenamefont {Devakul},\
  and\ \citenamefont {Fu}}]{zhang2021spintextured}%
  \BibitemOpen
  \bibfield  {author} {\bibinfo {author} {\bibfnamefont {Y.}~\bibnamefont
  {Zhang}}, \bibinfo {author} {\bibfnamefont {T.}~\bibnamefont {Devakul}},\
  and\ \bibinfo {author} {\bibfnamefont {L.}~\bibnamefont {Fu}},\ }\bibfield
  {title} {\bibinfo {title} {Spin-textured {{Chern}} bands in {{AB-stacked}}
  transition metal dichalcogenide bilayers},\ }\bibfield  {journal} {\bibinfo
  {journal} {PNAS}\ }\textbf {\bibinfo {volume} {118}},\ \href
  {https://doi.org/10.1073/pnas.2112673118} {10.1073/pnas.2112673118} (\bibinfo
  {year} {2021}{\natexlab{a}})\BibitemShut {NoStop}%
\bibitem [{\citenamefont {Devakul}\ and\ \citenamefont
  {Fu}(2021)}]{devakul2021quantum}%
  \BibitemOpen
  \bibfield  {author} {\bibinfo {author} {\bibfnamefont {T.}~\bibnamefont
  {Devakul}}\ and\ \bibinfo {author} {\bibfnamefont {L.}~\bibnamefont {Fu}},\
  }\bibfield  {title} {\bibinfo {title} {Quantum anomalous {{Hall}} effect from
  inverted charge transfer gap},\ }\href {http://arxiv.org/abs/2109.13909}
  {\bibfield  {journal} {\bibinfo  {journal} {arXiv:2109.13909}\ } (\bibinfo
  {year} {2021})}\BibitemShut {NoStop}%
\bibitem [{Het()}]{HeteroBilayer_SM}%
  \BibitemOpen
  \href@noop {} {\bibinfo {title} {See supplemental {{Material}}, which include
  \cite{wu2020quantum, fukui2005chern, monkhorst1976special}.}}\BibitemShut
  {Stop}%
\bibitem [{\citenamefont {Wu}\ and\ \citenamefont
  {Das~Sarma}(2020{\natexlab{a}})}]{wu2020quantum}%
  \BibitemOpen
  \bibfield  {author} {\bibinfo {author} {\bibfnamefont {F.}~\bibnamefont
  {Wu}}\ and\ \bibinfo {author} {\bibfnamefont {S.}~\bibnamefont {Das~Sarma}},\
  }\bibfield  {title} {\bibinfo {title} {Quantum geometry and stability of
  moir\'e flatband ferromagnetism},\ }\href
  {https://doi.org/10.1103/PhysRevB.102.165118} {\bibfield  {journal} {\bibinfo
   {journal} {Phys. Rev. B}\ }\textbf {\bibinfo {volume} {102}},\ \bibinfo
  {pages} {165118} (\bibinfo {year} {2020}{\natexlab{a}})}\BibitemShut
  {NoStop}%
\bibitem [{\citenamefont {Fukui}\ \emph {et~al.}(2005)\citenamefont {Fukui},
  \citenamefont {Hatsugai},\ and\ \citenamefont {Suzuki}}]{fukui2005chern}%
  \BibitemOpen
  \bibfield  {author} {\bibinfo {author} {\bibfnamefont {T.}~\bibnamefont
  {Fukui}}, \bibinfo {author} {\bibfnamefont {Y.}~\bibnamefont {Hatsugai}},\
  and\ \bibinfo {author} {\bibfnamefont {H.}~\bibnamefont {Suzuki}},\
  }\bibfield  {title} {\bibinfo {title} {Chern {{Numbers}} in {{Discretized
  Brillouin Zone}}: {{Efficient Method}} of {{Computing}} ({{Spin}}) {{Hall
  Conductances}}},\ }\href {https://doi.org/10.1143/JPSJ.74.1674} {\bibfield
  {journal} {\bibinfo  {journal} {J. Phys. Soc. Jpn.}\ }\textbf {\bibinfo
  {volume} {74}},\ \bibinfo {pages} {1674} (\bibinfo {year}
  {2005})}\BibitemShut {NoStop}%
\bibitem [{\citenamefont {Monkhorst}\ and\ \citenamefont
  {Pack}(1976)}]{monkhorst1976special}%
  \BibitemOpen
  \bibfield  {author} {\bibinfo {author} {\bibfnamefont {H.~J.}\ \bibnamefont
  {Monkhorst}}\ and\ \bibinfo {author} {\bibfnamefont {J.~D.}\ \bibnamefont
  {Pack}},\ }\bibfield  {title} {\bibinfo {title} {Special points for
  {{Brillouin-zone}} integrations},\ }\href
  {https://doi.org/10.1103/PhysRevB.13.5188} {\bibfield  {journal} {\bibinfo
  {journal} {Phys. Rev. B}\ }\textbf {\bibinfo {volume} {13}},\ \bibinfo
  {pages} {5188} (\bibinfo {year} {1976})}\BibitemShut {NoStop}%
\bibitem [{\citenamefont {Wu}\ \emph {et~al.}(2018{\natexlab{b}})\citenamefont
  {Wu}, \citenamefont {Lovorn}, \citenamefont {Tutuc},\ and\ \citenamefont
  {MacDonald}}]{wu2018hubbard}%
  \BibitemOpen
  \bibfield  {author} {\bibinfo {author} {\bibfnamefont {F.}~\bibnamefont
  {Wu}}, \bibinfo {author} {\bibfnamefont {T.}~\bibnamefont {Lovorn}}, \bibinfo
  {author} {\bibfnamefont {E.}~\bibnamefont {Tutuc}},\ and\ \bibinfo {author}
  {\bibfnamefont {A.~H.}\ \bibnamefont {MacDonald}},\ }\bibfield  {title}
  {\bibinfo {title} {Hubbard model physics in transition metal dichalcogenide
  moir\'e bands},\ }\href {https://doi.org/10.1103/PhysRevLett.121.026402}
  {\bibfield  {journal} {\bibinfo  {journal} {Phys. Rev. Lett.}\ }\textbf
  {\bibinfo {volume} {121}},\ \bibinfo {pages} {026402} (\bibinfo {year}
  {2018}{\natexlab{b}})}\BibitemShut {NoStop}%
\bibitem [{\citenamefont {Hu}\ and\ \citenamefont
  {MacDonald}(2021)}]{hu2021competing}%
  \BibitemOpen
  \bibfield  {author} {\bibinfo {author} {\bibfnamefont {N.~C.}\ \bibnamefont
  {Hu}}\ and\ \bibinfo {author} {\bibfnamefont {A.~H.}\ \bibnamefont
  {MacDonald}},\ }\bibfield  {title} {\bibinfo {title} {Competing magnetic
  states in transition metal dichalcogenide moir\'e materials},\ }\href
  {https://doi.org/10.1103/PhysRevB.104.214403} {\bibfield  {journal} {\bibinfo
   {journal} {Phys. Rev. B}\ }\textbf {\bibinfo {volume} {104}},\ \bibinfo
  {pages} {214403} (\bibinfo {year} {2021})}\BibitemShut {NoStop}%
\bibitem [{\citenamefont {Ahn}\ and\ \citenamefont
  {Das~Sarma}(2022)}]{ahn2022disorderinduced}%
  \BibitemOpen
  \bibfield  {author} {\bibinfo {author} {\bibfnamefont {S.}~\bibnamefont
  {Ahn}}\ and\ \bibinfo {author} {\bibfnamefont {S.}~\bibnamefont
  {Das~Sarma}},\ }\bibfield  {title} {\bibinfo {title} {Disorder-induced
  two-dimensional metal-insulator transition in moir\'e transition metal
  dichalcogenide multilayers},\ }\href
  {https://doi.org/10.1103/PhysRevB.105.115114} {\bibfield  {journal} {\bibinfo
   {journal} {Phys. Rev. B}\ }\textbf {\bibinfo {volume} {105}},\ \bibinfo
  {pages} {115114} (\bibinfo {year} {2022})}\BibitemShut {NoStop}%
\bibitem [{\citenamefont {Xu}\ \emph {et~al.}(2020)\citenamefont {Xu},
  \citenamefont {Liu}, \citenamefont {Rhodes}, \citenamefont {Watanabe},
  \citenamefont {Taniguchi}, \citenamefont {Hone}, \citenamefont {Elser},
  \citenamefont {Mak},\ and\ \citenamefont {Shan}}]{xu2020correlated}%
  \BibitemOpen
  \bibfield  {author} {\bibinfo {author} {\bibfnamefont {Y.}~\bibnamefont
  {Xu}}, \bibinfo {author} {\bibfnamefont {S.}~\bibnamefont {Liu}}, \bibinfo
  {author} {\bibfnamefont {D.~A.}\ \bibnamefont {Rhodes}}, \bibinfo {author}
  {\bibfnamefont {K.}~\bibnamefont {Watanabe}}, \bibinfo {author}
  {\bibfnamefont {T.}~\bibnamefont {Taniguchi}}, \bibinfo {author}
  {\bibfnamefont {J.}~\bibnamefont {Hone}}, \bibinfo {author} {\bibfnamefont
  {V.}~\bibnamefont {Elser}}, \bibinfo {author} {\bibfnamefont {K.~F.}\
  \bibnamefont {Mak}},\ and\ \bibinfo {author} {\bibfnamefont {J.}~\bibnamefont
  {Shan}},\ }\bibfield  {title} {\bibinfo {title} {Correlated insulating states
  at fractional fillings of moir\'e superlattices},\ }\href
  {https://doi.org/10.1038/s41586-020-2868-6} {\bibfield  {journal} {\bibinfo
  {journal} {Nature}\ }\textbf {\bibinfo {volume} {587}},\ \bibinfo {pages}
  {214} (\bibinfo {year} {2020})}\BibitemShut {NoStop}%
\bibitem [{\citenamefont {Li}\ \emph {et~al.}(2021{\natexlab{b}})\citenamefont
  {Li}, \citenamefont {Li}, \citenamefont {Regan}, \citenamefont {Wang},
  \citenamefont {Zhao}, \citenamefont {Kahn}, \citenamefont {Yumigeta},
  \citenamefont {Blei}, \citenamefont {Taniguchi}, \citenamefont {Watanabe},
  \citenamefont {Tongay}, \citenamefont {Zettl}, \citenamefont {Crommie},\ and\
  \citenamefont {Wang}}]{li2021imaging}%
  \BibitemOpen
  \bibfield  {author} {\bibinfo {author} {\bibfnamefont {H.}~\bibnamefont
  {Li}}, \bibinfo {author} {\bibfnamefont {S.}~\bibnamefont {Li}}, \bibinfo
  {author} {\bibfnamefont {E.~C.}\ \bibnamefont {Regan}}, \bibinfo {author}
  {\bibfnamefont {D.}~\bibnamefont {Wang}}, \bibinfo {author} {\bibfnamefont
  {W.}~\bibnamefont {Zhao}}, \bibinfo {author} {\bibfnamefont {S.}~\bibnamefont
  {Kahn}}, \bibinfo {author} {\bibfnamefont {K.}~\bibnamefont {Yumigeta}},
  \bibinfo {author} {\bibfnamefont {M.}~\bibnamefont {Blei}}, \bibinfo {author}
  {\bibfnamefont {T.}~\bibnamefont {Taniguchi}}, \bibinfo {author}
  {\bibfnamefont {K.}~\bibnamefont {Watanabe}}, \bibinfo {author}
  {\bibfnamefont {S.}~\bibnamefont {Tongay}}, \bibinfo {author} {\bibfnamefont
  {A.}~\bibnamefont {Zettl}}, \bibinfo {author} {\bibfnamefont {M.~F.}\
  \bibnamefont {Crommie}},\ and\ \bibinfo {author} {\bibfnamefont
  {F.}~\bibnamefont {Wang}},\ }\bibfield  {title} {\bibinfo {title} {Imaging
  two-dimensional generalized {{Wigner}} crystals},\ }\href
  {https://doi.org/10.1038/s41586-021-03874-9} {\bibfield  {journal} {\bibinfo
  {journal} {Nature}\ }\textbf {\bibinfo {volume} {597}},\ \bibinfo {pages}
  {650} (\bibinfo {year} {2021}{\natexlab{b}})}\BibitemShut {NoStop}%
\bibitem [{\citenamefont {Liu}\ \emph {et~al.}(2021)\citenamefont {Liu},
  \citenamefont {Taniguchi}, \citenamefont {Watanabe}, \citenamefont {Gabor},
  \citenamefont {Cui},\ and\ \citenamefont {Lui}}]{liu2021excitonic}%
  \BibitemOpen
  \bibfield  {author} {\bibinfo {author} {\bibfnamefont {E.}~\bibnamefont
  {Liu}}, \bibinfo {author} {\bibfnamefont {T.}~\bibnamefont {Taniguchi}},
  \bibinfo {author} {\bibfnamefont {K.}~\bibnamefont {Watanabe}}, \bibinfo
  {author} {\bibfnamefont {N.~M.}\ \bibnamefont {Gabor}}, \bibinfo {author}
  {\bibfnamefont {Y.-T.}\ \bibnamefont {Cui}},\ and\ \bibinfo {author}
  {\bibfnamefont {C.~H.}\ \bibnamefont {Lui}},\ }\bibfield  {title} {\bibinfo
  {title} {Excitonic and {{Valley-Polarization Signatures}} of {{Fractional
  Correlated Electronic Phases}} in a {${\mathrm{WSe}}_{2}/{\mathrm{WS}}_{2}$}
  {{Moir\'e Superlattice}}},\ }\href
  {https://doi.org/10.1103/PhysRevLett.127.037402} {\bibfield  {journal}
  {\bibinfo  {journal} {Phys. Rev. Lett.}\ }\textbf {\bibinfo {volume} {127}},\
  \bibinfo {pages} {037402} (\bibinfo {year} {2021})}\BibitemShut {NoStop}%
\bibitem [{\citenamefont {Regan}\ \emph {et~al.}(2020)\citenamefont {Regan},
  \citenamefont {Wang}, \citenamefont {Jin}, \citenamefont {Bakti~Utama},
  \citenamefont {Gao}, \citenamefont {Wei}, \citenamefont {Zhao}, \citenamefont
  {Zhao}, \citenamefont {Zhang}, \citenamefont {Yumigeta}, \citenamefont
  {Blei}, \citenamefont {Carlstr{\"o}m}, \citenamefont {Watanabe},
  \citenamefont {Taniguchi}, \citenamefont {Tongay}, \citenamefont {Crommie},
  \citenamefont {Zettl},\ and\ \citenamefont {Wang}}]{regan2020mott}%
  \BibitemOpen
  \bibfield  {author} {\bibinfo {author} {\bibfnamefont {E.~C.}\ \bibnamefont
  {Regan}}, \bibinfo {author} {\bibfnamefont {D.}~\bibnamefont {Wang}},
  \bibinfo {author} {\bibfnamefont {C.}~\bibnamefont {Jin}}, \bibinfo {author}
  {\bibfnamefont {M.~I.}\ \bibnamefont {Bakti~Utama}}, \bibinfo {author}
  {\bibfnamefont {B.}~\bibnamefont {Gao}}, \bibinfo {author} {\bibfnamefont
  {X.}~\bibnamefont {Wei}}, \bibinfo {author} {\bibfnamefont {S.}~\bibnamefont
  {Zhao}}, \bibinfo {author} {\bibfnamefont {W.}~\bibnamefont {Zhao}}, \bibinfo
  {author} {\bibfnamefont {Z.}~\bibnamefont {Zhang}}, \bibinfo {author}
  {\bibfnamefont {K.}~\bibnamefont {Yumigeta}}, \bibinfo {author}
  {\bibfnamefont {M.}~\bibnamefont {Blei}}, \bibinfo {author} {\bibfnamefont
  {J.~D.}\ \bibnamefont {Carlstr{\"o}m}}, \bibinfo {author} {\bibfnamefont
  {K.}~\bibnamefont {Watanabe}}, \bibinfo {author} {\bibfnamefont
  {T.}~\bibnamefont {Taniguchi}}, \bibinfo {author} {\bibfnamefont
  {S.}~\bibnamefont {Tongay}}, \bibinfo {author} {\bibfnamefont
  {M.}~\bibnamefont {Crommie}}, \bibinfo {author} {\bibfnamefont
  {A.}~\bibnamefont {Zettl}},\ and\ \bibinfo {author} {\bibfnamefont
  {F.}~\bibnamefont {Wang}},\ }\bibfield  {title} {\bibinfo {title} {Mott and
  generalized {{Wigner}} crystal states in {{WSe}}{$_2$}/{{WS}}{$_2$} moir\'e
  superlattices},\ }\href {https://doi.org/10.1038/s41586-020-2092-4}
  {\bibfield  {journal} {\bibinfo  {journal} {Nature}\ }\textbf {\bibinfo
  {volume} {579}},\ \bibinfo {pages} {359} (\bibinfo {year}
  {2020})}\BibitemShut {NoStop}%
\bibitem [{\citenamefont {Zhou}\ \emph {et~al.}(2021)\citenamefont {Zhou},
  \citenamefont {Sung}, \citenamefont {Brutschea}, \citenamefont {Esterlis},
  \citenamefont {Wang}, \citenamefont {Scuri}, \citenamefont {Gelly},
  \citenamefont {Heo}, \citenamefont {Taniguchi}, \citenamefont {Watanabe},
  \citenamefont {Zar{\'a}nd}, \citenamefont {Lukin}, \citenamefont {Kim},
  \citenamefont {Demler},\ and\ \citenamefont {Park}}]{zhou2021bilayer}%
  \BibitemOpen
  \bibfield  {author} {\bibinfo {author} {\bibfnamefont {Y.}~\bibnamefont
  {Zhou}}, \bibinfo {author} {\bibfnamefont {J.}~\bibnamefont {Sung}}, \bibinfo
  {author} {\bibfnamefont {E.}~\bibnamefont {Brutschea}}, \bibinfo {author}
  {\bibfnamefont {I.}~\bibnamefont {Esterlis}}, \bibinfo {author}
  {\bibfnamefont {Y.}~\bibnamefont {Wang}}, \bibinfo {author} {\bibfnamefont
  {G.}~\bibnamefont {Scuri}}, \bibinfo {author} {\bibfnamefont {R.~J.}\
  \bibnamefont {Gelly}}, \bibinfo {author} {\bibfnamefont {H.}~\bibnamefont
  {Heo}}, \bibinfo {author} {\bibfnamefont {T.}~\bibnamefont {Taniguchi}},
  \bibinfo {author} {\bibfnamefont {K.}~\bibnamefont {Watanabe}}, \bibinfo
  {author} {\bibfnamefont {G.}~\bibnamefont {Zar{\'a}nd}}, \bibinfo {author}
  {\bibfnamefont {M.~D.}\ \bibnamefont {Lukin}}, \bibinfo {author}
  {\bibfnamefont {P.}~\bibnamefont {Kim}}, \bibinfo {author} {\bibfnamefont
  {E.}~\bibnamefont {Demler}},\ and\ \bibinfo {author} {\bibfnamefont
  {H.}~\bibnamefont {Park}},\ }\bibfield  {title} {\bibinfo {title} {Bilayer
  {{Wigner}} crystals in a transition metal dichalcogenide heterostructure},\
  }\href {https://doi.org/10.1038/s41586-021-03560-w} {\bibfield  {journal}
  {\bibinfo  {journal} {Nature}\ }\textbf {\bibinfo {volume} {595}},\ \bibinfo
  {pages} {48} (\bibinfo {year} {2021})}\BibitemShut {NoStop}%
\bibitem [{\citenamefont {Pan}\ \emph {et~al.}(2020{\natexlab{b}})\citenamefont
  {Pan}, \citenamefont {Wu},\ and\ \citenamefont {Das~Sarma}}]{pan2020quantum}%
  \BibitemOpen
  \bibfield  {author} {\bibinfo {author} {\bibfnamefont {H.}~\bibnamefont
  {Pan}}, \bibinfo {author} {\bibfnamefont {F.}~\bibnamefont {Wu}},\ and\
  \bibinfo {author} {\bibfnamefont {S.}~\bibnamefont {Das~Sarma}},\ }\bibfield
  {title} {\bibinfo {title} {Quantum phase diagram of a {{Moir\'e-Hubbard}}
  model},\ }\href {https://doi.org/10.1103/PhysRevB.102.201104} {\bibfield
  {journal} {\bibinfo  {journal} {Phys. Rev. B}\ }\textbf {\bibinfo {volume}
  {102}},\ \bibinfo {pages} {201104} (\bibinfo {year}
  {2020}{\natexlab{b}})}\BibitemShut {NoStop}%
\bibitem [{\citenamefont {Wang}\ \emph {et~al.}(2020)\citenamefont {Wang},
  \citenamefont {Shih}, \citenamefont {Ghiotto}, \citenamefont {Xian},
  \citenamefont {Rhodes}, \citenamefont {Tan}, \citenamefont {Claassen},
  \citenamefont {Kennes}, \citenamefont {Bai}, \citenamefont {Kim},
  \citenamefont {Watanabe}, \citenamefont {Taniguchi}, \citenamefont {Zhu},
  \citenamefont {Hone}, \citenamefont {Rubio}, \citenamefont {Pasupathy},\ and\
  \citenamefont {Dean}}]{wang2020correlated}%
  \BibitemOpen
  \bibfield  {author} {\bibinfo {author} {\bibfnamefont {L.}~\bibnamefont
  {Wang}}, \bibinfo {author} {\bibfnamefont {E.-M.}\ \bibnamefont {Shih}},
  \bibinfo {author} {\bibfnamefont {A.}~\bibnamefont {Ghiotto}}, \bibinfo
  {author} {\bibfnamefont {L.}~\bibnamefont {Xian}}, \bibinfo {author}
  {\bibfnamefont {D.~A.}\ \bibnamefont {Rhodes}}, \bibinfo {author}
  {\bibfnamefont {C.}~\bibnamefont {Tan}}, \bibinfo {author} {\bibfnamefont
  {M.}~\bibnamefont {Claassen}}, \bibinfo {author} {\bibfnamefont {D.~M.}\
  \bibnamefont {Kennes}}, \bibinfo {author} {\bibfnamefont {Y.}~\bibnamefont
  {Bai}}, \bibinfo {author} {\bibfnamefont {B.}~\bibnamefont {Kim}}, \bibinfo
  {author} {\bibfnamefont {K.}~\bibnamefont {Watanabe}}, \bibinfo {author}
  {\bibfnamefont {T.}~\bibnamefont {Taniguchi}}, \bibinfo {author}
  {\bibfnamefont {X.}~\bibnamefont {Zhu}}, \bibinfo {author} {\bibfnamefont
  {J.}~\bibnamefont {Hone}}, \bibinfo {author} {\bibfnamefont {A.}~\bibnamefont
  {Rubio}}, \bibinfo {author} {\bibfnamefont {A.~N.}\ \bibnamefont
  {Pasupathy}},\ and\ \bibinfo {author} {\bibfnamefont {C.~R.}\ \bibnamefont
  {Dean}},\ }\bibfield  {title} {\bibinfo {title} {Correlated electronic phases
  in twisted bilayer transition metal dichalcogenides},\ }\href
  {https://doi.org/10.1038/s41563-020-0708-6} {\bibfield  {journal} {\bibinfo
  {journal} {Nature Materials}\ }\textbf {\bibinfo {volume} {19}},\ \bibinfo
  {pages} {861} (\bibinfo {year} {2020})}\BibitemShut {NoStop}%
\bibitem [{\citenamefont {Gu}\ \emph {et~al.}(2021)\citenamefont {Gu},
  \citenamefont {Ma}, \citenamefont {Liu}, \citenamefont {Watanabe},
  \citenamefont {Taniguchi}, \citenamefont {Hone}, \citenamefont {Shan},\ and\
  \citenamefont {Mak}}]{gu2021dipolar}%
  \BibitemOpen
  \bibfield  {author} {\bibinfo {author} {\bibfnamefont {J.}~\bibnamefont
  {Gu}}, \bibinfo {author} {\bibfnamefont {L.}~\bibnamefont {Ma}}, \bibinfo
  {author} {\bibfnamefont {S.}~\bibnamefont {Liu}}, \bibinfo {author}
  {\bibfnamefont {K.}~\bibnamefont {Watanabe}}, \bibinfo {author}
  {\bibfnamefont {T.}~\bibnamefont {Taniguchi}}, \bibinfo {author}
  {\bibfnamefont {J.~C.}\ \bibnamefont {Hone}}, \bibinfo {author}
  {\bibfnamefont {J.}~\bibnamefont {Shan}},\ and\ \bibinfo {author}
  {\bibfnamefont {K.~F.}\ \bibnamefont {Mak}},\ }\bibfield  {title} {\bibinfo
  {title} {Dipolar excitonic insulator in a moire lattice},\ }\href
  {http://arxiv.org/abs/2108.06588} {\bibfield  {journal} {\bibinfo  {journal}
  {arXiv:2108.06588}\ } (\bibinfo {year} {2021})}\BibitemShut {NoStop}%
\bibitem [{\citenamefont {Zhang}\ \emph
  {et~al.}(2021{\natexlab{b}})\citenamefont {Zhang}, \citenamefont {Regan},
  \citenamefont {Wang}, \citenamefont {Zhao}, \citenamefont {Wang},
  \citenamefont {Sayyad}, \citenamefont {Yumigeta}, \citenamefont {Watanabe},
  \citenamefont {Taniguchi}, \citenamefont {Tongay}, \citenamefont {Crommie},
  \citenamefont {Zettl}, \citenamefont {Zaletel},\ and\ \citenamefont
  {Wang}}]{zhang2021correlated}%
  \BibitemOpen
  \bibfield  {author} {\bibinfo {author} {\bibfnamefont {Z.}~\bibnamefont
  {Zhang}}, \bibinfo {author} {\bibfnamefont {E.~C.}\ \bibnamefont {Regan}},
  \bibinfo {author} {\bibfnamefont {D.}~\bibnamefont {Wang}}, \bibinfo {author}
  {\bibfnamefont {W.}~\bibnamefont {Zhao}}, \bibinfo {author} {\bibfnamefont
  {S.}~\bibnamefont {Wang}}, \bibinfo {author} {\bibfnamefont {M.}~\bibnamefont
  {Sayyad}}, \bibinfo {author} {\bibfnamefont {K.}~\bibnamefont {Yumigeta}},
  \bibinfo {author} {\bibfnamefont {K.}~\bibnamefont {Watanabe}}, \bibinfo
  {author} {\bibfnamefont {T.}~\bibnamefont {Taniguchi}}, \bibinfo {author}
  {\bibfnamefont {S.}~\bibnamefont {Tongay}}, \bibinfo {author} {\bibfnamefont
  {M.}~\bibnamefont {Crommie}}, \bibinfo {author} {\bibfnamefont
  {A.}~\bibnamefont {Zettl}}, \bibinfo {author} {\bibfnamefont {M.~P.}\
  \bibnamefont {Zaletel}},\ and\ \bibinfo {author} {\bibfnamefont
  {F.}~\bibnamefont {Wang}},\ }\bibfield  {title} {\bibinfo {title} {Correlated
  interlayer exciton insulator in double layers of monolayer {{WSe}}{$_2$} and
  moir\'e {{WS}}{$_2$}/{{WSe}}{$_2$}},\ }\href
  {http://arxiv.org/abs/2108.07131} {\bibfield  {journal} {\bibinfo  {journal}
  {arXiv:2108.07131}\ } (\bibinfo {year} {2021}{\natexlab{b}})}\BibitemShut
  {NoStop}%
\bibitem [{\citenamefont {Pan}\ and\ \citenamefont
  {Das~Sarma}(2021)}]{pan2021interactiondriven}%
  \BibitemOpen
  \bibfield  {author} {\bibinfo {author} {\bibfnamefont {H.}~\bibnamefont
  {Pan}}\ and\ \bibinfo {author} {\bibfnamefont {S.}~\bibnamefont
  {Das~Sarma}},\ }\bibfield  {title} {\bibinfo {title} {Interaction-{{Driven
  Filling-Induced Metal-Insulator Transitions}} in {{2D Moir\'e Lattices}}},\
  }\href {https://doi.org/10.1103/PhysRevLett.127.096802} {\bibfield  {journal}
  {\bibinfo  {journal} {Phys. Rev. Lett.}\ }\textbf {\bibinfo {volume} {127}},\
  \bibinfo {pages} {096802} (\bibinfo {year} {2021})}\BibitemShut {NoStop}%
\bibitem [{\citenamefont {Pan}\ and\ \citenamefont
  {Das~Sarma}(2022)}]{pan2022interaction}%
  \BibitemOpen
  \bibfield  {author} {\bibinfo {author} {\bibfnamefont {H.}~\bibnamefont
  {Pan}}\ and\ \bibinfo {author} {\bibfnamefont {S.}~\bibnamefont
  {Das~Sarma}},\ }\bibfield  {title} {\bibinfo {title} {Interaction range and
  temperature dependence of symmetry breaking in strongly correlated
  two-dimensional moir\'e transition metal dichalcogenide bilayers},\ }\href
  {https://doi.org/10.1103/PhysRevB.105.041109} {\bibfield  {journal} {\bibinfo
   {journal} {Phys. Rev. B}\ }\textbf {\bibinfo {volume} {105}},\ \bibinfo
  {pages} {041109} (\bibinfo {year} {2022})}\BibitemShut {NoStop}%
\bibitem [{\citenamefont {Wu}\ and\ \citenamefont
  {Das~Sarma}(2020{\natexlab{b}})}]{wu2020collective}%
  \BibitemOpen
  \bibfield  {author} {\bibinfo {author} {\bibfnamefont {F.}~\bibnamefont
  {Wu}}\ and\ \bibinfo {author} {\bibfnamefont {S.}~\bibnamefont {Das~Sarma}},\
  }\bibfield  {title} {\bibinfo {title} {Collective {{Excitations}} of
  {{Quantum Anomalous Hall Ferromagnets}} in {{Twisted Bilayer Graphene}}},\
  }\href {https://doi.org/10.1103/PhysRevLett.124.046403} {\bibfield  {journal}
  {\bibinfo  {journal} {Phys. Rev. Lett.}\ }\textbf {\bibinfo {volume} {124}},\
  \bibinfo {pages} {046403} (\bibinfo {year} {2020}{\natexlab{b}})}\BibitemShut
  {NoStop}%
\bibitem [{\citenamefont {Tse}\ and\ \citenamefont
  {MacDonald}(2010)}]{tse2010giant}%
  \BibitemOpen
  \bibfield  {author} {\bibinfo {author} {\bibfnamefont {W.-K.}\ \bibnamefont
  {Tse}}\ and\ \bibinfo {author} {\bibfnamefont {A.~H.}\ \bibnamefont
  {MacDonald}},\ }\bibfield  {title} {\bibinfo {title} {Giant {{Magneto-Optical
  Kerr Effect}} and {{Universal Faraday Effect}} in {{Thin-Film Topological
  Insulators}}},\ }\href {https://doi.org/10.1103/PhysRevLett.105.057401}
  {\bibfield  {journal} {\bibinfo  {journal} {Phys. Rev. Lett.}\ }\textbf
  {\bibinfo {volume} {105}},\ \bibinfo {pages} {057401} (\bibinfo {year}
  {2010})}\BibitemShut {NoStop}%
\end{thebibliography}%

\end{document}